\documentclass[conference]{IEEEtran}
\ifCLASSINFOpdf
\usepackage[pdftex]{graphicx}
\else
\fi
\usepackage{algorithm}
\usepackage[noend]{algpseudocode}

\usepackage{glossaries} 

\usepackage[numbers]{natbib}

\usepackage{tabularx}
\usepackage{xcolor}

\usepackage{booktabs}

\usepackage{multirow}
\usepackage{amsthm} 

\theoremstyle{definition}

\usepackage{array}

\usepackage[tight,footnotesize]{subfigure}
\newglossary[slg]{symbolslist}{syi}{syg}{Notationlist} 

\newglossarystyle{notationlong}{%
	\setglossarystyle{long}
		{\end{longtable}}%

}

\begin{document}
	%
	\title{Mobile Link Prediction: Automated Creation and Crowd-sourced Validation of Knowledge Graphs}

	\author{\IEEEauthorblockN{Mark C. Ballandies}
		\IEEEauthorblockA{Computational Social Science \\
			ETH Zurich\\
			Stampfenbachstrasse 48, 8092 Zurich\\ 
			Email: bmark@ethz.ch}
		\and
		\IEEEauthorblockN{Evangelos Pournaras}
		\IEEEauthorblockA{School of Computing \\
			University of Leeds\\
			Leeds LS2 9JT, UK\\ 
			E.Pournaras@leeds.ac.uk}}
	
	
	%


	\maketitle

	\begin{abstract}
			Building trustworthy knowledge graphs for cyber-physical social systems (CPSS) is a challenge. In particular, current approaches relying on human experts have limited scalability, while automated approaches are often not accountable to users resulting in knowledge graphs of questionable quality. This paper introduces a novel pervasive knowledge graph builder that brings together automation, experts' and crowd-sourced citizens' knowledge. The knowledge graph grows via automated link predictions using genetic programming that are validated by humans for improving transparency and calibrating accuracy. The knowledge graph builder is designed for pervasive devices such as smartphones and preserves privacy by localizing all computations. The accuracy, practicality, and usability of the knowledge graph builder is evaluated in a real-world social experiment that involves a smartphone implementation and a Smart City application scenario. The proposed knowledge graph building methodology outperforms the baseline method in terms of accuracy while demonstrating its efficient calculations on smartphones and the feasibility of the pervasive human supervision process in terms of high interactions throughput. These findings promise new opportunities to crowd-source and operate pervasive reasoning systems for cyber-physical social systems in Smart Cities.

	\end{abstract}
	

	%
	\IEEEpeerreviewmaketitle
	
	\glsaddall
	
	\newglossaryentry{ao_gloss}
	{
		name=Abstract Object,
		description={netti. term for a \textit{concept} in the Ontology}
	}
	\newglossaryentry{ar_gloss}
	{
		name=Abstract Relationship,
		description={netti. term for a \textit{property} in the Ontology}
	}
	\newglossaryentry{co_gloss}
	{
		name=Concrete Object,
		description={netti. term for a \textit{instance} in the Ontology}
	}
	\newglossaryentry{thing_gloss}{name={Thing},description={a resource or relationship in the netti. ontology}}
	
	\newacronym{ao}{AO}{Abstract Object}
	\newacronym{ar}{AR}{Abstract Relationship}
	\newacronym{co}{CO}{Concrete Object}

	
	\newglossaryentry{symb:n_gloss}{name=\ensuremath{N},
		description={number of different node types},
		type=symbolslist}
	\newglossaryentry{symb:m_gloss}{name=\ensuremath{M},
		description={number of different link types},
		type=symbolslist}
	\newglossaryentry{symb:vi_gloss}{name=\ensuremath{V_i},
		description={set of nodes of the same type $i$},
		type=symbolslist}
	\newglossaryentry{symb:realized_link_gloss}{name=\ensuremath{(u,v,j)},
		description={realized link: $u,v \in V, j\in \{0,..M\}$ },
		type=symbolslist}
	\newglossaryentry{symb:realized_link_of_type_gloss}{name=\ensuremath{E_j},
		description={set of all realized links of type $j$ },
		type=symbolslist}

\section{Introduction}
\label{sec:introduction}

Mobile cyber-physical systems involve humans utilizing mobile services in their social contexts. This inclusion of human actors extends the classical cyber-physical systems paradigm \cite{chakraborty2016automotive} to cyber-physical-\textit{social} systems (CPSS) \cite{dautov2018data}.  These systems integrate both, social and physical systems by intelligent human-machine interactions in cyber-physical space \cite{zhang2018cyber}.

Knowledge graphs store information in a graph structure that are often utilized in these CPSS to improve services such as route navigation~\cite{dudas2009onalin}, health recommendations~\cite{hu2016personal} \cite{wiesner2010adapting} or question answering \cite{hao2017end}. In particular, knowledge graphs improve the performance of learning algorithms at predicting unobserved relationships between entities in an application domain~\cite{caragea2009ontology, cui2019infer}. Nevertheless, manually building knowledge graphs may be impractical and unscalable~\cite{alani2003automatic}. Hence systems utilizing link prediction methods are proposed to automate the building of knowledge graphs~\cite{wang2013boosting}.

These CPSSs are designed either explicitly or implicitly for values such as usability \cite{schafer2017towards}, autonomy \cite{dussell1999position}, or privacy \cite{asikis2020value}. The sucessful implementation of these values into CPSSs can determine their adoption by humans \cite{zhou2011effect} \cite{belanger2008trust, nahavandi2017trusted} and thus should be explicitly accounted for in the design phase~\cite{friedman1997software}.

Hence, this work applies a value-sensitive design methodology~\cite{van2008moral, van2010use, friedman1997software, hill1991autonomy, friedman2003human} that explicitly considers values such as privacy and accountability to design a CPSS in the form of a knowledge graph builder that constructs a knowledge graph by adding links.  By utilizing a novel link prediction methodology, the knowledge graph building is automated. 
In particular, users are assisted to identify missing relationships in a knowledge graph via a link prediction method. By following a privacy-by-design approach, both the knowledge graph as well as the link prediction method are deployed locally on users' mobile phones without access from a third-party. The automated knowledge graph building remains accountable-by-design to humans by letting users supervise the accuracy of recommendations via accepting or rejecting recommended links. Moreover, as this feedback is then in turn utilized to train the link prediction method, users can control the calibration of their machine intelligence. This value-sensitive design approach builds a trustworthy domain-specific knowledge graph about users' reality that can improve services provided by CPSS such as privacy-preserving recommenders. 

The contributions of this work are the following:
\begin{itemize}
	\item An automated knowledge graph builder for CPSSs that is accountable via human-supervision, preserves the privacy of its users and runs locally on smart phones. In particular, the novel approach connects expert knowledge, automation and crowd-sourcing to collaboratively build a trustworthy and personalized knowledge graph.
	\item The extension of an existing link prediction methodology \cite{bliss2014evolutionary} with structural semantic and temporal information.
	\item Extended and novel similarity metrics that measure the probability of link formation between two nodes of a knowledge graph.
	\item Identification of dominant metric ensembles that guide link prediction in knowledge graphs in a smart city application scenario.
\end{itemize}

This paper is organized as follows: In Section \ref{sec:related_work}, content-based recommenders and knowledge graph building via link prediction are discussed. A data model for knowledge graphs and its applications for digital assistance is introduced in Section \ref{sec:data_model}, while the automated and privacy-preserving knowledge graph builder is then outlined in Section \ref{sec:decision_support_system}. Thereafter, Section \ref{sec:methodology} illustrates the methodology of the conducted experiment and Section \ref{sec:evaluation} presents the evaluation. Section \ref{sec:summary_of_findings} summarizes the findings and Section \ref{sec:conclusion} draws a conclusion and gives an outlook on future work.

\section{Background and literature review}
\label{sec:related_work}


Intelligent CPSSs in the form of recommenders are studied to sort through information and to make personalized recommendations to individual users~\cite{pazzani1999framework}.
Two types of methods are applied in these recommender systems~ \cite{van2000using}: User-based collaborative and content-based filtering.
The former
is often not privacy-preserving as it relies on collecting sensitive information from users~\cite{basilico2004unifying,isinkaye2015recommendation,friedman2015privacy}. In contrast, the latter relies on informative content descriptors \cite{basilico2004unifying} in the form of a common and transparent information source that can be constructed by expert knowledge \cite{asikis2020value}, crowd-sourced information \cite{carlier2011combining, goeau2011visual}, or automation \cite{ferman2002content}. Often this approach does not rely on sensitive user information and thus can better preserve their privacy. In particular, this approach optimizes recommendations by matching users' preferences (e.g. watched products) with product information. A novel approach in content-based recommender systems that follows a value-sensitive design improves product recommendations while shopping by matching local user personalization with a centrally maintained information source in the form of a knowledge graph \cite{asikis2020value},  which has been shown to improve recommender systems \cite{yu2014personalized}. By performing this matching on the users' phone, the accuracy of recommendations is improved while users' privacy is preserved.
Nevertheless, as knowledge graphs are often static and     incomplete \cite{shi2016discriminative, melo2017approach}, this approach misses the opportunity to improve recommendations by letting users build the utilized knowledge graph  \cite{cui2019infer}.



In general, such knowledge graph building can either be performed by (i)~human experts (ii)~crowd-sourcing, or (iii)~automation \cite{nickel2015review}. Utilizing human experts results in highly accurate knowledge graphs, but lacks scalability due to the limited available human resources \cite{wang2019research}. Crowd-sourcing information scales better but may result in less accurate knowledge graphs \cite{west2014knowledge}. Additionally, the scalability of that approach, though increased, can also become saturated as the slow down of Wikipedia growth indicates \cite{suh2009singularity}. Hence, automating knowledge graph building is promising to increase its scalability. Nevertheless, it is a challenge to determine the accuracy of automatically constructed knowledge graphs which reduces their trustworthiness \cite{nickel2015review}. 


Two tasks in knowledge graph building are identified~\cite{paulheim2017knowledge}: Knowledge graph completion and error detection. The former focuses on adding new instances (e.g. links) to the knowledge graph whereas the latter identifies and removes erroneous information from the knowledge graph. 
To tackle these challenges with automation and thus scale up the knowledge graph building, two types of methods are utilized~\cite{nickel2015review}: latent feature and graph feature-based methods.
Latent feature-based methods often lack the capability to account for new entities entering the knowledge graph as those are not considered in the latent feature calculations~\cite{amador2019ontology}. Moreover, these methods utilize the whole knowledge graph in their calculation which can result in limited scalability and privacy concerns~\cite{cui2019infer}. 
In contrast, graph-based methods utilize the knowledge graph directly to calculate features. Three types of methods are identified that utilize graph-based information~\cite{nickel2015review}: Similarity measures, rule mining and inductive logic programming, and path rank algorithms.

Similarity-based methods is the most commonly used approach in link prediction \cite{chen2016fast}. In this approach, a score is assigned to new candidate links, and the top-k links with the highest score are recommended \cite{haghani2019systemic}. These algorithms require no domain knowledge to compute the similarity scores \cite{lichtenwalter2010new} and can identify homophily patterns in knowledge graphs~\cite{haghani2019systemic}.
Depending on the structural information utilized in the calculations, similarity measures can be clustered into three groups \cite{martinez2016survey}:  local, quasi-local, and global similarity metrics. When compared to the computations of global metrics, local similarity metrics computations are more efficient and parallelizable but are restricted on distance-two nodes (neighbors of neighbors)~\cite{martinez2016survey}. Quasi-local similarity metrics are less efficient when compared to local metrics but can in contrast to those assign similarity scores to further apart nodes~\cite{martinez2016survey}.

Recently, knowledge graphs grew to networks consisting of thousands of different object and link types \citep{cao2016link}. These networks are often incomplete and change dynamically, which makes mining and analysis challenging \cite{friemel:dynamics_of_social_networks,wang2015link}.
In particular, link prediction in such networks has to model topological as well as temporal and semantic influences between various types of relationships and to identify the underlying mechanisms that drive the formation of new relationships \cite{davis2013supervised}.
Extensive reviews of link prediction are outlined in Wang et al. \cite{wang2015link} for social networks, in Shi~et~al.~\cite{shi2017survey} for networks with more than one relationship type and in Martinez~et~al.~\cite{martinez2016survey} for complex networks. 
In the following, an overview of the link prediction literature that is important for this work is illustrated.

Tylenda et. al \cite{tylenda2009towards} find that temporal information about changes in knowledge graphs are a dominant feature in link prediction. This has been confirmed by Yang et al. \cite{yang2012predicting} by introducing supervised and unsupervised methods for link prediction in knowledge graphs. Moreover, the authors introduce the multi-relational influence propagation metric for heterogeneous networks.     
Likewise, other researchers developed measures and algorithms utilizing the ontology of knowledge graphs. For instance, Maedche et al.  \cite{maedche2002clustering} introduce relation and taxonomy similarity metrics to measure the similarity between any two objects in a knowledge graph by analyzing ontological information. It was shown that these measures perform well for cluster analysis \cite{grimnes2008instance}.
Likewise, Opuszko et. al \cite{opuszko2012classification} predict links between actors using ontology-based similarity measures. They show that including these measures can improve the prediction performance.
Nonetheless, they also show, that their results are often not easily interpretable and that it is not obvious how to weight the different measures when combined. In particular, this often requires domain knowledge and manual effort~\cite{ma2019jointly}. This is confirmed by Brando et. al \cite{brandao2013using} who state that the weighting of different metrics is a grant challenge in the context of link prediction.

Bliss et. al \cite{bliss2014evolutionary} address this problem by utilizing a genetic algorithm for link prediction \citep{bliss2014evolutionary}, which adjusts the weights of the similarity measures by optimization. They use local similarity measures to estimate the likelihood of an unobserved link existence. 
The strength of this approach compared to other link prediction algorithms is that it neither requires the assumption of network classes nor prior knowledge about the analyzed knowledge graph as the weights are calculated by the optimization strategy \citep{ozcan2019multivariate}. 
It is shown that this approach produces comparable results to other link prediction approaches while enabling researchers to analyze the networks driving mechanisms. In particular, the change of weights of different similarity measures during a period of time or for different networks can be analyzed \citep{bliss2014evolutionary}. Often neither a single metric is dominant for predictions \cite{vani2015investigating} nor is the combination of metrics stable over different application domains. Thus identification of dominant metric weights in novel application domains is required. 
Nevertheless, Bliss et. al \cite{bliss2014evolutionary} primarily focus on topological information of a homogenous network \cite{ozcan2019multivariate} and thus they neither investigate the performance of temporal and ontology-based similarity metrics nor the applicability of their method on heterogeneous\footnote{Network consisting of more than one entity type, as defined in Section \ref{sec:linked_graph}} or multi-dimensional\footnote{Networks consisting of more than one relationship type, as defined in Section \ref{sec:linked_graph}} networks consisting of a multitude of link and node types.

In summary, current approaches in recommender systems often lack privacy-preservation and rely on a centralized completion of global knowledge graphs which does not scale well when applied in a smartphone setting. The identification of dominant combinations of similarity metrics for an application domain is challenging and the approach of Bliss et al. \cite{bliss2014evolutionary} does not consider temporal and ontological similarity metrics which might improve prediction accuracy. 

In order to solve these identified gaps, this work extends the method of Bliss et. al \cite{bliss2014evolutionary} with temporal, ontology-based, local, and quasi-local similarity measures that are applied to a multi-dimensional and heterogeneous knowledge graph to identify the dominant similarity metrics in a smart city application scenario. 
The method is then utilized in a human-supervised and privacy-preserving knowledge graph builder to enable users to build a personalized knowledge graph in a given application scenario that collaboratively combines experts' knowledge, crowd-sourced information, and automation.

\section{Automated knowledge graph building via link prediction}
\label{sec:data_model}

This article focuses on knowledge graphs that are modeled as an ontology. Such a structure enhances machines/ algorithms capability to analyze and interpret information \cite{maedche2002clustering}. In the following the concept of an ontology is introduced (Section \ref{sec:ontology}) and its data model is defined (Section \ref{sec:linked_graph}). Moreover, applications of knowledge graphs are illustrated (Section \ref{sec:application_ontology}).

\subsection{Ontology} 
\label{sec:ontology}

\begin{figure}[!t]
	\centering
	\includegraphics[width=1.0\columnwidth]{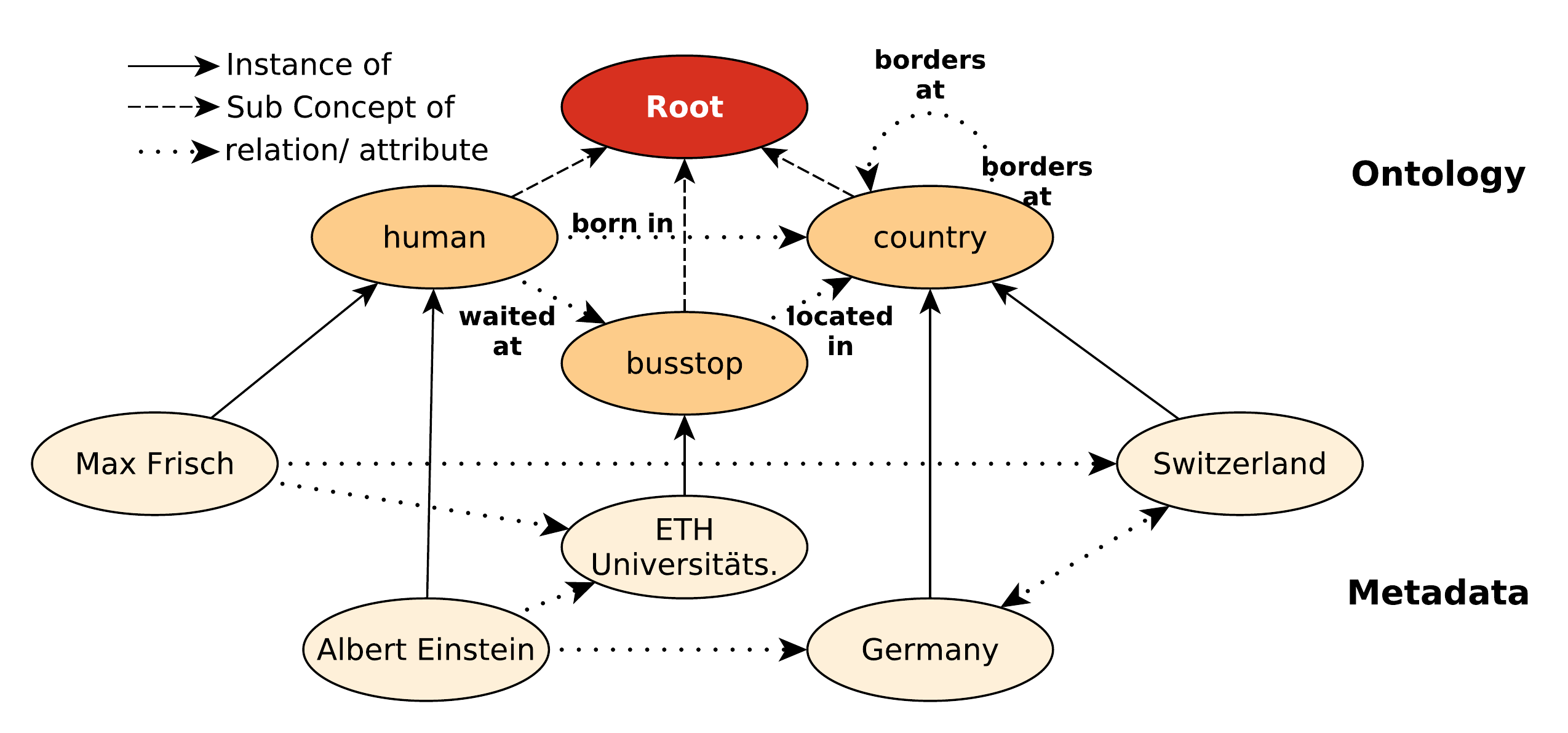}
	\caption{Example knowledge graph consisting of an ontology and metadata structure that construct a knowledge graph.
	}
	\label{fig:ontology}
\end{figure}

Ontologies formally define types, properties, and relationships between entities that are applied to a concrete domain and enable the construction of knowledge graphs \cite{amador2019ontology}.
Ontologies and all related concepts are rigorously defined in Maedche et. al \citep{maedche2002clustering}. In the following, those terms and terminologies are introduced that are relevant for this work. An example of an ontology and an instance of it - a metadata structure, which together construct a knowledge graph, are depicted in Figure~\ref{fig:ontology}. 
An ontology consists of concepts (e.g. \textit{human} in Figure \ref{fig:ontology}) and relation identifiers (e.g. \textit{waited at} in Figure~\ref{fig:ontology}). The concepts are structured in a concept hierarchy (e.g. human is a \textit{sub concept} of root). Concepts are instantiated by instances (e.g. Albert Einstein is an instance of a human in Figure \ref{fig:ontology}). Moreover, a \textit{concrete relationship} is an instantiation of a relationship between two instances (e.g. Albert Einstein- \textit{born in} - Germany is a concrete relationship in Figure \ref{fig:ontology}).

\begin{table}
	\caption{Symbols and expressions utilized in this work.}
	\label{tab:link_pred_comparison}
	\centering
	\resizebox{1.0\columnwidth}{!}{
		
		\begin{tabular}{ ll}
			\toprule
			\textbf{Symbol} &\textbf{Explanation} \\
			\midrule
			$V_i$ & set of nodes of concept $i$ \\
			$V$ & $(V_1 \cup V_2 \cup ... \cup V_N)$\\
			$E_j$ & set of realized triplets of link type $j$ \\
			$E$ & $(E_1 \cup E_2 \cup ... \cup E_M)$\\
			$G(V,E)$ & graph consisting of nodes in $V$ and links in $E$\\
			$ \Gamma(u,j)$ &neighborhood of $u$ for link type $j: \{v \in V | \exists (u,v,j) \in E_j \}$ \\
			$\Gamma(u)$& neighborhood of $u: \cup_{j\in M} \Gamma(u,j)$\\
			$k_u$ & degree of node $u$ \\
			$P_n$ & path of length $n$ between $u$ and $v$\\
			$J(u) $&  set of all links in which $u$ is subject: \\
			&$\{ j \in M| \exists v \in V: (u,v,j) \in E \}$ \\
			$J(u,v)$& set of realized links in which $u$ is the subject  \\
			& and $v$ the object node: $\{ j \in M| u,v \in V: (u,v,j) \in E \}$ \\
			$E(j)$& set of node pairs that are connected via a relation identifier $j$.\\ 
			& $\{(u,v)| \exists (u,v,j) \in E \}$ \\
			$N(j)$ &  set of subject nodes:$ \{u \in V| \exists (u,v,j) \in E \}$ \\
			
			\bottomrule
		\end{tabular}
	}
	
\end{table}

\subsection{Directed graphs as a data model for knowledge representations}
\label{sec:linked_graph}

As depicted in Figure \ref{fig:ontology}, an instance of a knowledge graph can be modeled as a graph. In the following necessary definitions are given.

Let $i \in \{1,...,N\}$ be a concept (i.e. human, country, bus stop) and $N$ the number of different concepts, then $V_i$ is the set of nodes of the same concept $i$ and $V = (V_1 \cup V_2 \cup ... \cup V_N)$ is the set of all nodes. An instance of a concept $i$ can be denoted as $u_i$ or as $u\in V_i$.
A relation identifier~$j~\in~\{1,..,M\} := S_M$ (e.g. waited at, born, etc.) can connect two nodes $u,v$    ; $M$ being the number of different relation identifiers. A concrete relationship is denoted as a triplet $(u,v,j)$, where $u,v \in V$ and $j$ denote the relation identifier. These links are directed, where $u$ is the subject and $v$ the object. The set of realized triplets of link type $j$ is then denoted as $\gls{symb:realized_link_of_type_gloss}$ and $E=(E_1 \cup E_2 \cup ... \cup E_M)$ is the set of all realized links. 
$e \in E$ is referred to as concrete relationship or link in the following.

The graph or network with nodes in $V$ and links in $E$ is then denoted as $G(V,E)$.
If $N>1$ the graph $G$ is called heterogeneous and if $M > 1$ the graph is called multi-dimensional.    
On such a multi-dimensional and heterogeneous network a similarity measure $s_i$ can be defined rigorously as in Chen et. al \cite{chen:similarity_metric}. 
Each $s_i$ measures how similar two nodes $u,v$ are. 
The linear combination of such metrics
\begin{equation}
\label{eq:linear_combined_similarity_metrics}
\begin{aligned}
s(u,v) =& \sum_{i} a_i s_i(u,v), \\
&\sum_i a_i =1.
\end{aligned}
\end{equation} 
is also a similarity metric \cite{chen:similarity_metric}. 
A similarity measure can be normalized onto the range $[0,1]$.

\subsection{Automated knowledge graph building via link prediction}
\label{sec:application_ontology}

In this work, link prediction is utilized to automate the completion of a knowledge graph by predicting links between existing instances.
In the following, the link prediction problem is introduced (Section \ref{sec:problem_formulation}). A method is then proposed in Section \ref{sec:decision_support_system}.

\subsubsection{Problem formulation}
\label{sec:problem_formulation}

The link prediction task on a multidimensional and heterogeneous graph $G$ (Section \ref{sec:linked_graph}) can be stated as follow:
Let $G_p(V,E_p) \subset G(V,E)$ be a sub graph such that $E_p \subset E$. The task of link prediction is to identify those $j \in E$ that are currently not observed in $E_p$.   
Let $u,v \in V$, $j \in S_M$ and $s_1: s_1(V,V,S_M) \rightarrow [0,1]$ and $s_2: s_2(V,V) \rightarrow [0,1]$ being normalized similarity measures. A similarity measure models the probability of an unobserved link to be established between the nodes. Two types of predictions are considered in this work:

\begin{itemize}
	\item \textbf{Existence prediction} is utilized in one-dimensional knowledge graphs ($M=1$) and predicts if any link exists between the two nodes $u$ and $v$ ($\exists j \in S_M s.t. (u,v,j) \in E \land (u,v,j) \notin E_p $), thus without specifying the relation identifier. The probability of such a link formation is defined as the normalized similarity measure $s_2(u,v)$ between the two nodes.
	\item \textbf{Semantic prediction: } is utilized in multi-dimensional knowledge graphs ($M>1$) and predicts what relation identifier connects the two nodes $u$ and $v$. The probability of a link $j \in S_M$ to be formed is defined as the similarity measure $s_1(u,v,j)$ between these two nodes along the candidate link.
\end{itemize}

This work distinguishes between these two types of predictions because of computational considerations: Heterogeneous and multi-dimensional knowledge graphs have complex dependency structures \cite{davis2013supervised}. 
Computing the semantic type of a link between two nodes requires distinguishing the formation mechanism for each link type \cite{davis2013supervised} which is computationally costly. Hence this work predicts the semantic type of relationship only between nodes where it is known that already a link exists. Existence prediction is utilized for nodes that are not connected yet.

\section{A human-supervised and privacy-preserving knowledge graph builder}
\label{sec:decision_support_system}

In this section, the knowledge-graph builder is introduced. 
It performs link prediction (Section \ref{sec:problem_formulation}) on a heterogeneous and multi-dimensional knowledge graph 
(Section \ref{sec:linked_graph}) utilizing an optimization mechanism in form of genetic programming (Section \ref{sec:genetic_algorithm}) to optimize the weights of various similarity metrics (Section \ref{sec:utilized_metrics}). These weighted metrics then facilitate the recommendation of missing links (Section \ref{sec:problem_formulation}) in a knowledge graph (Section \ref{sec:ontology}). Moreover, the builder is supervised by the user and preserves their privacy.

In the following, a background on genetic programming is given (Section \ref{sec:genetic_algorithm}) before the link prediction method is introduced (Section \ref{sec:algorithm}). Finally, the knowledge graph builder is illustrated (Section \ref{sec:decision_support_system}).

\subsection{Background: genetic programming}
\label{sec:genetic_algorithm}

Genetic programming is an optimization method that is utilized in symbolic regression to identify underlying functions to given data points that explain their dependencies \cite{koza:genetic}. Compared to other optimization strategies, genetic programming provides solutions for large, poorly defined search spaces that are high-dimensional, multi-modal, and noisy \cite{sastry2019survey}.
Due to its flexibility in adjusting to diverse problems \cite{espejo2009survey}, genetic programming has gained an increased interest in diverse research communities such as software improvement \cite{petke2017genetic}, image processing~\cite{khan2019recent}, production scheduling~\cite{nguyen2017genetic}, and machine learning \cite{agapitos2019survey}. 

\subsection{Link prediction method}
\label{sec:algorithm}
The method is an extension of the link prediction algorithm found in Bliss et. al \cite{bliss2014evolutionary} with temporal and ontology-based metrics added:
The core idea of the prediction algorithm is to measure topological, temporal and semantic similarity metrics~$s_i$ between a target node $u$ and candidate node $v$ on a multi-dimensional and heterogeneous graph $G$ and then to predict based on a weighted combination of these metrics if a link $j$ between these nodes exists (existence prediction, Section~ \ref{sec:problem_formulation}), resp. what type of link should be formed (semantic prediction, Section \ref{sec:problem_formulation}).

The link prediction method consists of two main steps and is depicted in Algorithm \ref{eq:netti_algorithm}. 
In the first step, the weights $a_i$ are obtained via genetic programming (Section \ref{sec:genetic_algorithm}). 
Then, Equation \ref{eq:linear_combined_similarity_metrics} is utilized to predict which links are unobserved in a given candidate set that was obtained by a baseline heuristic.    
In both steps, it is distinguished between semantic and existence prediction, as formulated in Section~\ref{sec:problem_formulation}. 

\begin{algorithm}
	\caption{Link Prediction Algorithm}\label{eq:netti_algorithm}
	\begin{algorithmic}[1]
		\Procedure{predict}{$u$}\Comment{obtain link existence or type information}
		\State $\vec{a} \gets \text{getWeights}()$ \Comment{see section \ref{sec:obtaining_weights}}
		\State $r \gets $ predictType($u,a$) or predictExistence($u,a$) \Comment{see section \ref{sec:prediction}}
		\State \textbf{return} $\vec{r}$\Comment{List of links (semantic prediction) or nodes (existence prediction) with assigned similarity values as calculated by Equation \ref{eq:linear_combined_similarity_metrics}.}
		\EndProcedure
	\end{algorithmic}
\end{algorithm}

\subsubsection{Training - Obtaining weights $a_i$}
\label{sec:obtaining_weights}
Algorithm \ref{eq:obtain_weights} depicts the weight calculation algorithm: It first calculates a training set and then uses this training set as an input for the genetic programming algorithm, which after termination returns the weight vector.

\begin{algorithm}
	\caption{Obtain Weights}\label{eq:obtain_weights}
	\begin{algorithmic}[1]
		\Procedure{getWeights}{ }\Comment{obtain weights for similarity measures}
		\State $\text{List$<$TrainingInstance$> l$ } \gets 
		\text{getTrainingInstances()}$
		\State $\vec{a} \gets \text{calcGenetic($l$)}$ 
		\State \textbf{return} $\vec{a}$
		\EndProcedure
	\end{algorithmic}
\end{algorithm}

The training set generation algorithm creates training sets consisting of positive and negative training instances. In link existence prediction, an instance consists of: $(u,v,\{0,1\})$, $u,v \in V$. In case of semantic prediction it has the following form: $(u,v,j,\{0,1\})$, $u,v \in V$, $j \in M$.
For both types, $1$ indicates that a link (of type $j$) exists between node $u$ and $v$ and $0$ that no link exists.
The training set consists of $50 \%$ positive (1) and $50 \%$ negative (0) instances. Depending on the available information, the algorithm operates in two modes for the generation of negative instances:
(i) Knowledge of non-existent links: It is known that specific links of type $j$ are not existent between some nodes $u,v \in V$. 
This often requires manual work but is considered as the gold standard in method evaluation \cite{paulheim2017knowledge}. (ii) \textit{No} knowledge of non-existing links: It is assumed that all \textit{unobserved} links in the network are non-existent. Due to the incompleteness of knowledge graphs \cite{shi2016discriminative, melo2017approach}, this approach is considered as the silver standard in evaluation \cite{paulheim2017knowledge}.

The details of how genetic programming is implemented in this work can be found in Appendix \ref{ap:training}.

\begin{figure*}[]
	\includegraphics[width=1\textwidth]{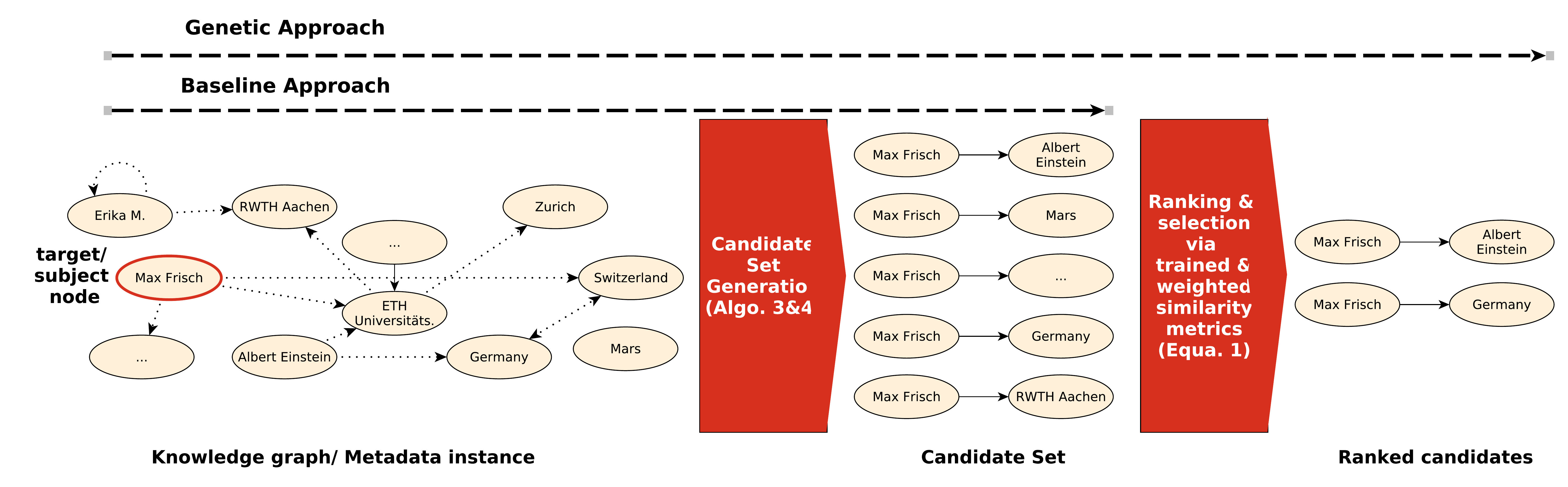}
	\caption{Example illustrating the baseline and genetic link prediction methods: For the given target node Max Frisch, the baseline methods (Algorithms \ref{eq:candidate_set_existence} and \ref{eq:candidate_set_type}) calculate a candidate set of unobserved existing links. The genetic link prediction method utilizes these candidate sets and creates ranked suggestions via the trained and weighted similarity metrics (Equation \ref{eq:linear_combined_similarity_metrics}).}
	\label{fig:transparent_link_prediction}
\end{figure*}

\subsubsection{Prediction}
\label{sec:prediction} 

\begin{algorithm}
	\caption{Existence baseline method}\label{eq:candidate_set_existence}
	\begin{algorithmic}[1]
		
		\Procedure{getExistanceCandidate}{$u,N$b}\Comment{Existence Candidate Set}
		\State $t\gets \text{getNeighbors}(u)$
		
		\State $s$ $\gets$ getNeighborsForEach($t$)
		\State $s$.remove($t$)
		\State $result$.add(random($s,N/2$))
		\State $s \gets $ getAllNodes()
		\State $s$.remove($t$)
		\State $result$.add(random($s,N/2$))
		\State \textbf{return} $result$\Comment{Candidate set of nodes}
		\EndProcedure
	\end{algorithmic}
\end{algorithm}

Figure \ref{fig:transparent_link_prediction} illustrates the utilized link prediction method:
In a first step, for a given target node $u$ (e.g. Max Frisch in Figure \ref{fig:transparent_link_prediction}) a candidate set $c$ is calculated by a baseline method. This method returns a set consisting of candidate nodes $v_i$ (existence prediction) 
or of candidate node-relationship pairs ($v_i,j$) (semantic prediction), $j\in S_M$.
Then, in a second step, Equation \ref{eq:linear_combined_similarity_metrics} is applied on the candidate set to obtain a similarity score for each candidate $(u,v_i)$, respectively $(u,v_i,j)$. The set is ordered based on these scores and the top entries are utilized for link prediction. In the evaluation (Section \ref{sec:evaluation}), the accuracy of these predictions are compared to the accuracy of taking randomly instances from the baseline's candidate set.

Two baseline methods are utilized for comparison, one for the existence and one for the semantic prediction.
The existence baseline method (Algorithm~\ref{eq:candidate_set_existence}) for a target node $u$ considers both, exploitation and exploration of the existing knowledge graph: In order to exploit topological information, $50 \%$ of the candidate set consists of neighbors of existing neighbors of $u$ that are not already connected to $u$. The other half is constructed by exploring the remaining knowledge graph and thus to include not connected nodes randomly with an equal probability. 

Because instances in knowledge graphs are often linked by more than one relation identifier, the semantic baseline method (Algorithm \ref{eq:candidate_set_type}) exclusively exploits topological information: The candidate set is constructed by the neighbors of $u$, which are included in the candidate set with an equal probability. Hence, in contrast to the existence baseline, the full knowledge graph is not explored to decrease the computational complexity as outlined in Section \ref{sec:problem_formulation}. Each of the selected nodes is then accompanied with possible relationship identifiers of links that still can be formed between the selected node and~$u$.

\begin{algorithm}
	\caption{Semantic baseline method}\label{eq:candidate_set_type}
	\begin{algorithmic}[1]
		
		\Procedure{getSemanticCandidate}{$u,N$}\Comment{Type Candidate Set generation}
		\State $s\gets \text{getNeighbors}(u)$
		\For{Node c : s}\Comment{for each candidate}
		\State rel $\gets$ chooseNonExistingRelationship($u,c$)
		\State $l$.add($c$,rel)
		\If{$l$.size()$>N$}
		\State break
		\EndIf
		\EndFor\label{euclidendwhile}
		\State \textbf{return} $l$\Comment{Candidate set of pairs (node,link)}
		\EndProcedure
	\end{algorithmic}
\end{algorithm}

\subsubsection{Utilized similarity metrics}
\label{sec:utilized_metrics}
The utilized metrics are clustered in three groups, characterized by their applicability in existence prediction, semantic prediction or both predictions. Moreover, the metrics are normalized to take values in the interval~$[0,1]$. 
In the following, additional notation to illustrate the metrics is introduced (Section \ref{sec:additional_notation}) before the utilized metrics are illustrated in greater detail (Section \ref{sec:metric_description}).

\paragraph{Additional notation}
\label{sec:additional_notation}
Besides the notation introduced in Section \ref{sec:linked_graph}, the following is required to define the utilized similarity metrics:
The neighborhood of $u$ for link type $j$ is defined as
$ \Gamma(u,j) = \{v \in V | \exists (u,v,j) \in E_j \}$. The neighborhood of $u$ is defined as $\Gamma(u)= \cup_{j\in M} \Gamma(u,j)$. In addition, $k_u$ is the degree of node~$u$.  
A path of length $n$ between $u$ and $v$ is denoted as $P_n$. $J(u) = \{ j \in M| \exists v \in V: (u,v,j) \in E \}$ is the set of all links in which $u$ is the subject node. $J(u,v) = \{ j \in M| u,v \in V: (u,v,j) \in E \}$ is the set of realized links in which $u$ is the subject and $v$ the object node.
Moreover, $E(j) = \{(u,v)| \exists (u,v,j) \in E \}$ is the set of node pairs that are connected via a relation identifier $j$ and $N(j) = \{u \in V| \exists (u,v,j) \in E \}$ is the set of subject nodes.

\begin{table*}
	\caption{Utilized metrics in the knowledge builder.}
	\label{tab:classical_metrics}
	\resizebox{1.0\textwidth}{!}{
		
		\begin{tabular}{ lllcclc}
			\toprule
			\textbf{ID} &\textbf{ Metric name}& \textbf{\begin{tabular}[c]{@{}c@{}}Equation \\  \end{tabular}} & \multicolumn{2}{c}{\textbf{Prediction}}&\textbf{Type}& \textbf{\begin{tabular}[c]{@{}c@{}}Normalized/ \\  \end{tabular} }\\
			& &  &Existence&Semantic& &\textbf{Novel }\\
			\midrule
			1 &Jaccard Index (J) \cite{jaccard1901bulletin}  & $ \frac{|\Gamma (u) \cap \Gamma(v)|}{|\Gamma(u) \cup \Gamma(v)|}$    &X&X&  topological  &\\

			2&Adamic Adar (AA) \cite{adamic2003friends} & 
			$ \Big(\sum_{z\in \Gamma (u) \cap \Gamma(v)} \frac{1}{log(|\Gamma(z)|)} \Big)$ $\frac{1}{|\Gamma(u)\cap \Gamma(v)|\frac{1}{log(2)}}$
			& X&  X& topological&X\\
			
			3& Resource Allocation (R)  \cite{zhou2009predicting} &   $  \Big( \sum_{z \in \Gamma(u) \cap \Gamma(v)} \frac{1}{|\Gamma(z)|} \Big)$ $\frac{1}{|\Gamma(u)\cap \Gamma(v)|\frac{1}{2}}$

			&X&X&  topological&X  \\ 
			
			4 & Hub promoted (Hp) \cite{ravasz2002hierarchical}&  $\frac{|\Gamma(u) \cap \Gamma(v)|}{max \{ k_u,k_v\}}$  & X  & X& topological& \\
			
			5&Hub depressed (Hd) \cite{ravasz2002hierarchical} &$\frac{|\Gamma(u) \cap \Gamma(v)|}{max \{ k_u,k_v\}}$ &X&  X& topological& \\
			
			6&Leicht-Holme-New. (L) \cite{leicht2006vertex} &$ \frac{|\Gamma(u) \cap \Gamma(v)|}{k_u k_v}$ & X&  X& topological&\\
			
			7&Salton (Sa)  \cite{mcgill1983introduction} & $ \frac{|\Gamma(u) \cap \Gamma(v)|}{\sqrt{k_u k_v}}$    &X&X&  topological & \\ 
			
			8&Sorenson index (So) \cite{sorensen1948method}&  $\frac{2 |\Gamma(u) \cap \Gamma(v)|}{k_u + k_v}$  & X  & X& topological& \\
			
			9&Shortest path (SP)&$max \Big(0, 1- \frac{\underset{n}{min}P_n(u,v)-1}{5}\Big) $ &X&  X&topological&X \\
			
			10&Time score (TS) \cite{munasinghe2012time} &$\sum_{c_i \in \Gamma(u) \cap \Gamma(v)} \frac{H_m^i \beta^{k_i}}{|t_1^i - t_2^i|+1}$ & X&  X&time&X\\
			11&Euler Time Metric (ET) & $e^{- \text{lastlink}(v)/d},
			d    = \text{discounting factor}$ & X & X &time & \\
			
			12&Focci distance (FD)\cite{jahanbakhsh2012predicting}&         $\underset{j\in J(u)\cap J(v)}{max}\underset{z\in \Gamma(u,j)\cap \Gamma(v,j)}{max} \frac{1}{|\Gamma(z,\text{inverse}(j))|}$

			&X&& topological, semantic&X\\
			
			13&Conditional Prob. (CP) \cite{yang2012predicting} & $ \underset{j\in M\setminus i}{max} \frac{| E(i) \cap E(j)| }{|E(i)|}$    & &X&semantic&   \\ 
			
			14&Taxonomy Sim. (OR) \cite{maedche2002clustering}&$\frac{\sum_{a \in A_s(P,I_1)} max  \{ sim(a,b)|b \in A_s(P,I_2)\}}{|A_s(P,I_1)|}$&X&&semantic&\\
			
			15&Relational Sim. (RS) \cite{maedche2002clustering}& 
			$\big(\sum_{p\in P_{co-I}}OR(I_1,I_2,p) + $ $ \sum_{p\in P_{co-O}} OR(I_1,I_2,p)\big)\frac{1}{|P_{co-I}+|P_{co-O}|}$ 
			&X&&semantic,  topological& \\

			16&    AR Relation (ARR)   & $ \frac{J(u) \cap J(v)}{J(u)}$    &X&&  semantic &X  \\

			17&AO Relation (AOR)    &$\frac{|\text{type}(u)== \text{type}(\Gamma(v))|}{|\Gamma(v)|}$ & X&  & semantic & X\\
			
			18&AO Relation Reversed (AORR)&   AOR(v,u)  & X&& semantic & X\\
			
			19&AO Relation Combined (AORC)& $\frac{AOR(u,v)+AORR(u,v)}{2}$  & X   &&semantic& X\\
			
			20&Node Dimension Conn. (NDC)  \cite{rossetti2011scalable} & $ \frac{|\{ u\in V | \exists v \in V: (u,v,i) \in E\}|}{|V|}$    &&X& topological   & \\ 
			
			21&Edge Dimension Conn. (EDC) \cite{rossetti2011scalable}&  $\frac{|\{(u,v,i)\in E | u, v \in V\}|}{|E|}$  &    & X&topological & \\
			22 & Multi Relational Link propagation \cite{yang2012predicting}&$score(v)\beta \frac{weight(v,u,i)}{degree(v,i)}+score(v)\beta \sum_{j\neq i}^K\big(\sigma(i,j) \frac{weight(v,u,j)}{degree(v,j)}/ (|E(v,u)|-1)\big)$&&X&topological, semantic\\
			
			\bottomrule
		\end{tabular}
	}
	
\end{table*}

\paragraph{Metric description}
\label{sec:metric_description}
Table \ref{tab:classical_metrics} depicts the 27 utilized metrics.
As illustrated in Section \ref{sec:problem_formulation}, this article distinguishes between semantic and existence prediction. Not all metrics can be utilized in both of these types of predictions (Columns 4 and 5 in Table \ref{tab:classical_metrics}). 19 metrics are topological metrics that utilize the metadata structure of the knowledge graph (Section \ref{sec:ontology}), two are time-based and 14 utilize semantic information of the knowledge graph by using its ontology (Column 6 in Table \ref{tab:classical_metrics}). 
Four metrics are introduced in this paper (ID 16-19 in Table \ref{tab:classical_metrics}), two metrics found in literature are modified such that they are normalized to take values in the range $[0,1]$ (ID $2, 3$ in Table \ref{tab:classical_metrics}) and three metrics are adjusted such that they can be utilized with the information model ((ID $9-11$) in Table \ref{tab:classical_metrics}). These novel and modified metrics are illustrated in greater detail in the Appendix.

\subsection{knowledge graph builder for cyber-physical systems}
\label{sec:knowledge_graph_builder}

\begin{figure}[!htb]
	\includegraphics[width=\columnwidth]{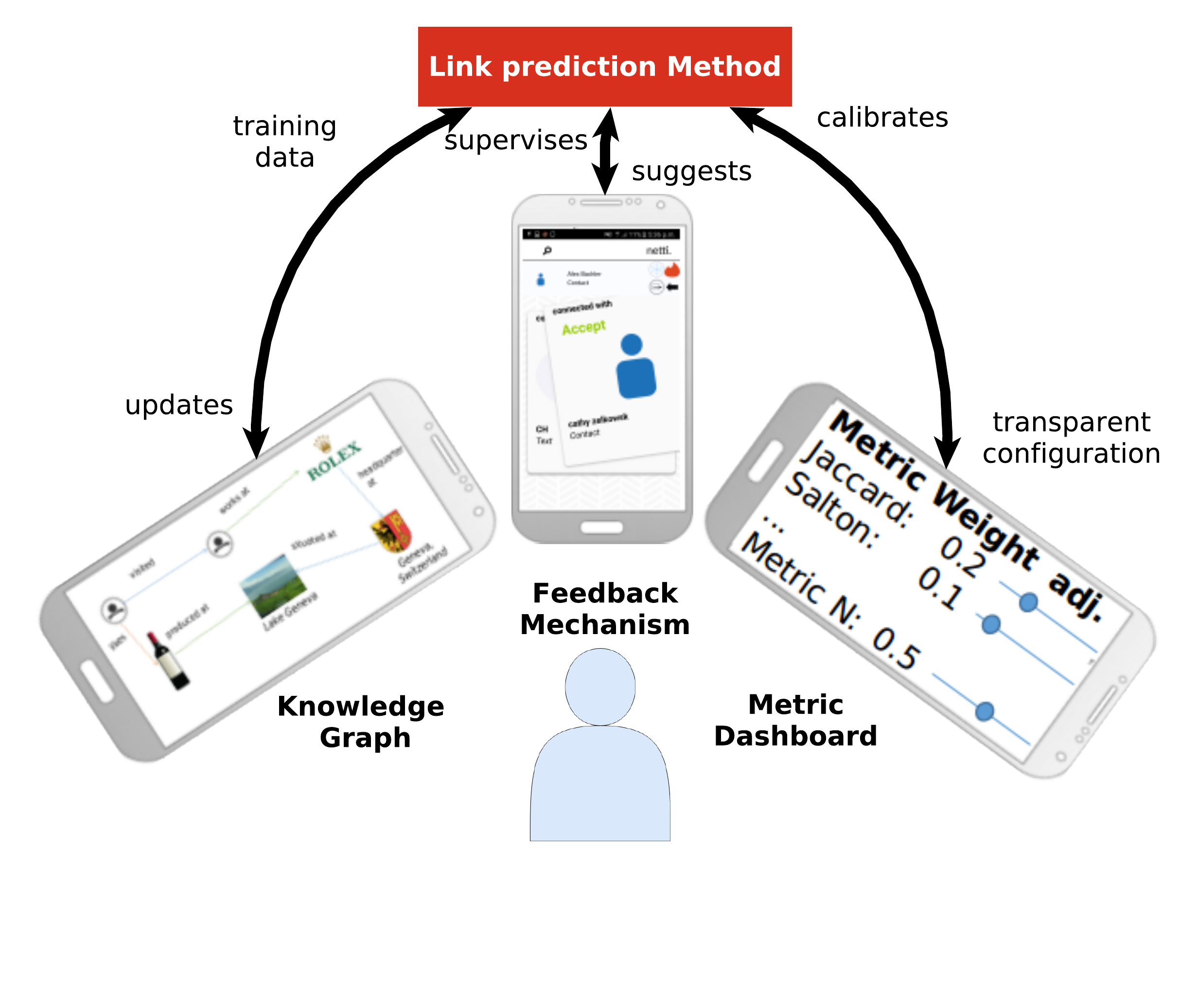}
	\caption{Privacy-preserving and human-supervised knowledge graph builder for crowd-sourcing knowledge graphs in cyber-physical systems. It consists of a link prediction method, knowledge graph visualisations, a feedback mechanism, and metric dashboard.}
	\label{fig:system_layout}
\end{figure}

The knowledge graph builder consists of the link prediction method, a feedback mechanism called tinder view, knowledge graph visualizations, and a metric weight dashboard. They are integrated into users' daily life by deploying the builder as a mobile application on users' phones, as depicted in Figure \ref{fig:system_layout}. 

\begin{figure}[!htb]
	\includegraphics[width=\columnwidth]{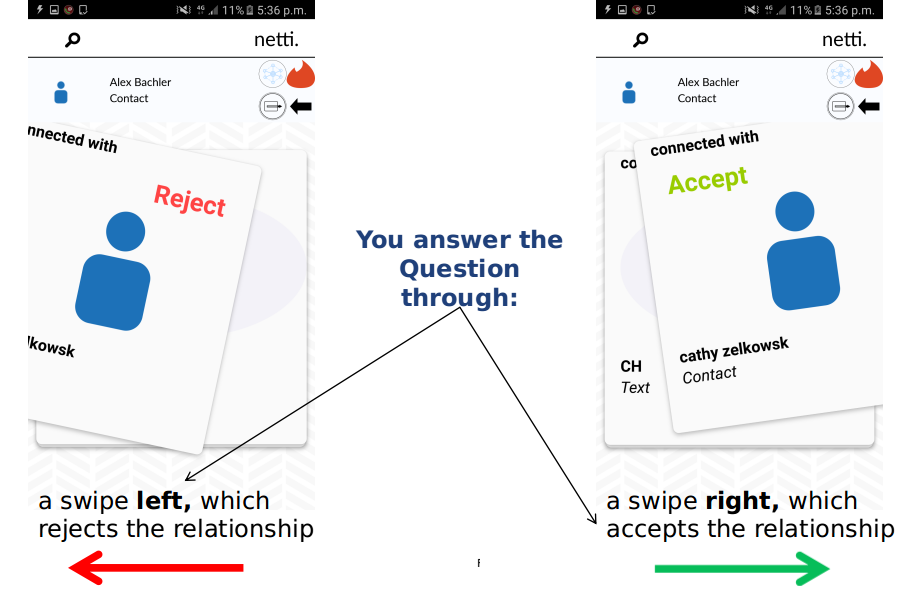}
	\caption{Illustration of the feedback mechanism, as presented to the experiment participants. Links can be rejected or accepted by swipes. In the example, the user is asked if the contact Alex Bachler is connected with Cathy Zelkowsk. Any instance of the knowledge base such as bus stops or cities could be presented instead of a contact.}
	\label{fig:tinder}
\end{figure}

The user can supervise the link prediction method by giving feedback via the tinder view as depicted in Figure \ref{fig:tinder}. In this way the completion of the knowledge graph happens in a supervised way: links are recommended by the algorithm and final decisions for their acceptance or rejection are performed by the users which is a necessary condition for users' autonomy \cite{crisp1987persuasive}. 
Based on this supervision (information about existing and non-existing links), weights of the link prediction methods are updated, as illustrated in Section~\ref{sec:algorithm}. In particular, by utilizing this evaluation strategy that is considered as the \textit{gold standard} in validation \cite{paulheim2017knowledge}, the builder continuously collects information about non-existing links that are utilized in the training set construction of the genetic programming.    The obtained weights are presented to the users in the metric dashboard. Moreover, users can adjust the metric weights in the dashboard and thus control the link prediction mechanism. Moreover, they can learn how the algorithm is configured and thus reason about link predictions.
Finally, as the data and algorithms are deployed locally without the requirement to communicate with a centralized server, the privacy of the users is preserved. In particular, no information about the metric weights or the knowledge graph are revealed to third parties.

\section{Experiment Methodology}
\label{sec:methodology}
The knowledge graph builder is evaluated by a social experiment. Additionally, the system is utilized to investigate the performance of various similarity metrics to predict links. In the following, the methodology of the experiment is illustrated. In particular, hypotheses are introduced (Section \ref{sec:hypotheses}), the data schema is illustrated (Section \ref{sec:schema_utilized}) and the experiment execution is illustrated (Section \ref{sec:experiment_methodology}).

\subsection{Hypotheses and Operationalisation}
\label{sec:hypotheses}

\subsubsection{Knowledge graph builder}
\label{sec:hyp_decision_support}

The usability of the knowledge graph builder and its accuracy in predicting unobserved links in a knowledge graph is investigated by the following hypotheses:
\paragraph{The knowledge graph builder improves the accuracy of link prediction compared to the baseline heuristic}
The knowledge graph builder utilizes a genetic programming approach to estimate the weights of various similarity measures (Table \ref{tab:classical_metrics}). These weights are then utilized to improve the accuracy in link prediction of the baseline method (Figure \ref{fig:transparent_link_prediction} and Algorithms \ref{eq:candidate_set_existence} - \ref{eq:candidate_set_type}).

Both methods are evaluated in the following way:
Users rate link suggestions (true/ false) via the feedback mechanism (Section \ref{sec:knowledge_graph_builder}) of the knowledge graph builder. One-third of these suggested links are drawn randomly from the baselines candidate set and two-third are taken from the highest-ranked results of the link prediction method (Figure \ref{fig:transparent_link_prediction}). The accuracy in the form of true positives is then evaluated.

\paragraph{The knowledge graph builder is usable measured in terms of user interaction}
As reasoned in Section \ref{sec:introduction}, the success of CPSS is dependent on its usability for humans. 
The usability of the knowledge graph builder is measured by analyzing the frequency with which users utilize the feedback mechanism of the knowledge graph builder to train the link prediction method.

\subsubsection{Metric weights}
\label{sec:hyp_metric_weights}

The dominance of different metric weights is investigated by the following hypothesis.

\paragraph{A combination of metrics compared to a single metric increases the accuracy of link predictions}
It is known from link prediction in energy grids that not all metrics show the same performance. Moreover, often the combination of several metrics outperforms a single metric \cite{mei2011power}. Hence, likewise, it is assumed that also in knowledge graph completion a combination of metrics increases the accuracy of predictions compared to single metrics.
The hypothesis is evaluated by analyzing the final metric weights obtained from genetic programming. Assuming that genetic programming maximizes accuracy, a single metric is dominant if its weight is close to one and those of all other metrics are zero.

\paragraph{Semantic and temporal metrics improve the link prediction performance}
Leveraging semantic and temporal information increase link prediction performance compared to a scenario that utilizes only topological information (Section \ref{sec:related_work}).    
This is evaluated by analyzing the metric weights obtained by the genetic programming algorithm. In particular, the weights of temporal and semantic metrics are compared to those of topological metrics.

\subsection{Knowledge graph instantiation: Model and schemas}
\label{sec:schema_utilized}

\begin{figure*}[!t]
	\centering
	\includegraphics[width=1.0\textwidth]{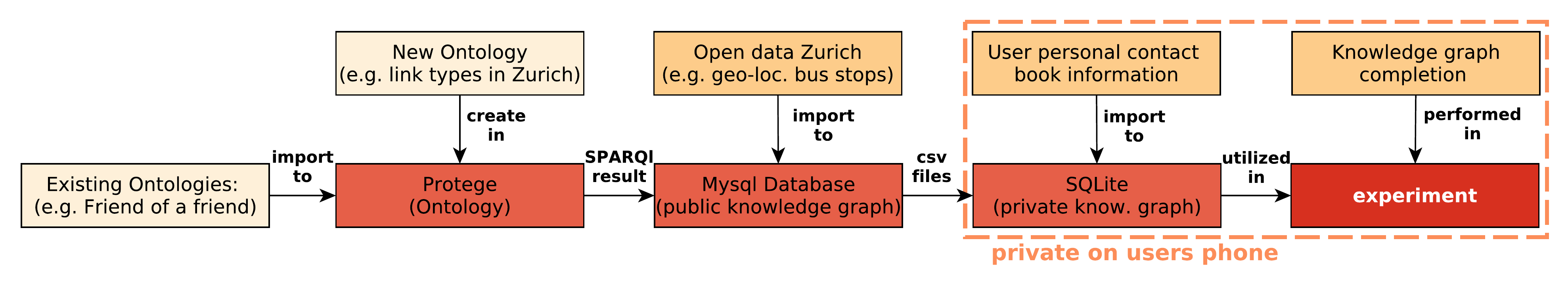}
	\caption{Instantiation flow of knowledge graph utilized in the mobile application for the experiment.            }
	\label{fig:data_flow}
\end{figure*}

The data model of the knowledge graph is a directed graph, as illustrated in Section \ref{sec:linked_graph}.
Figure~\ref{fig:data_flow} depicts the instantiation flow of the data model:
In a first step existing ontologies such as friends of a friend\footnote{Ontology that defines people related terms suitable for storing generalized user profile data, as well as social friendship relations \cite{friedman2015privacy}: http://www.foaf-project.org/ (last accessed: May 2020).} are merged in Protege\footnote{A free, open-source ontology editor and framework for building intelligent systems that uses the owl schema and is developed by the University of Stanford: http://protege.stanford.edu/.} and extended with relation identifiers and concepts illustrating a city. In particular, in order to support users completing their existing knowledge graph within the cyber-physical system of a smart city, relation identifiers (e.g. the visit relation), and concepts (e.g. bus stops) illustrating a city are added. 

In order to automate the process of storing information on an android mobile phone that utilizes SQLite\footnote{C-language library implementing a SQL database: https://www.sqlite.org/index.html (last accessed: May 2020).}, a relational data schema in MySQL workbench is created that is populated with the data from Protege. This data is extended with information from Open data Zurich illustrating tram and bus stops in Zurich (e.g., the geo-locations).

Finally, this data is exported to users' phones where the data is extended with personal contact book information of each user resulting in a personalized knowledge graph illustrating the users' social contacts in the city of Zurich. This approach preserves a user's privacy as all personal information is stored locally on users' phones.

\subsection{Setup}
\label{sec:experiment_methodology}

\begin{figure}[!htb]
	\includegraphics[width=\columnwidth]{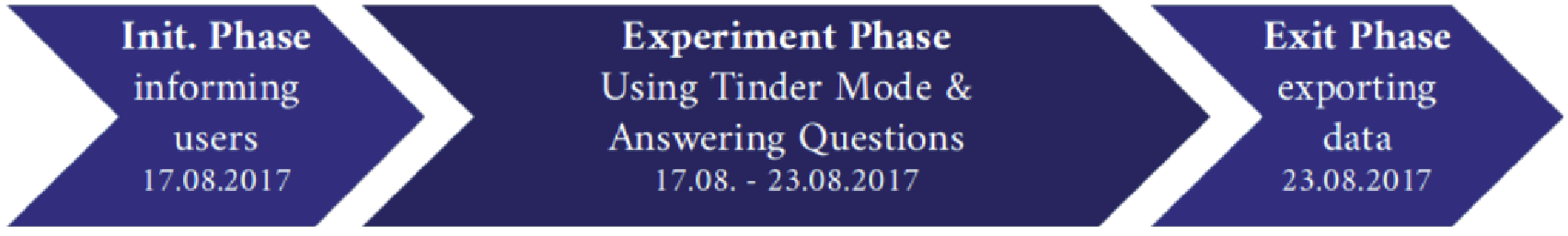}
	\caption{Three phases of the conducted experiment in which the users participated.}
	\label{fig:experiment_phases}
\end{figure}
\label{sec:process}
The experiment has an execution time of one week (17.08.2017 - 23.08.2017) and consists of three phases as depicted in Figure \ref{fig:experiment_phases}. The eleven participants are recruited by convenience sampling. In the initialization phase, users obtain a welcome email that contains detailed experiment instructions and which can be found in the Supplementary material. In the experiment phase, the users utilize the feedback mechanism and knowledge graph view of the knowledge builder (Figure \ref{fig:system_layout}) to complete their knowledge graph and to supervise the link prediction method. In the exit phase, users export their knowledge graph via an export button and send it by mail to the instructors of the experiment. During the export process, all personal data of the users are anonymized. 

\section{Experimental Evaluation}
\label{sec:evaluation}

By investigating the Hypotheses of Section \ref{sec:hypotheses}, both, the accuracy and usability of the knowledge builder (Section \ref{sec:eval_knowledge_builder}) as well as the capability of metrics to recommended unobserved links are analyzed (Section \ref{sec:eval_metric_weights}) in the following sections.

\subsection{Knowledge graph builder}
\label{sec:eval_knowledge_builder}

\begin{table}
	\caption{True and false positives of the genetic and baseline prediction methods.}
	\label{tab:link_pred_comparison}
	\centering
	\begin{tabular}{ lll}
		\toprule
		positive &\textbf{genetic} & \textbf{baseline}\\
		\midrule
		true & 0.2788 & 0.1220 \\
		false & 0.7212 & 0.8780 \\
		\bottomrule
	\end{tabular}

\end{table}
Table \ref{tab:link_pred_comparison} depicts the true (TP) and false (FP) positives of the genetic link prediction method compared to the baseline. The accuracy of the method is $27.9 \%$\footnote{In contrast to typical application scenarios in which recommender search spaces are small (e.g. types of pasta in a supermarket), the search space of the experiment is large consisting of every possible link between any two nodes in the knowledge graph. Hence,  this larger search space size could explain the lower TP probability of the applied method when compared to the performance of recommender systems in other typical scenarios. Also, a cold start of the algorithm is applied which initially can lower the TP probability and which could be analyzed in future work by extending the study period of the experiment.} and of the baseline is $12.20 \%$.

\begin{table}
	\caption{The table depicts the amount of feedback a user provided via the feedback mechanism of the knowledge graph builder, the true positives of the genetic approach, the true positives of the baseline method and the fraction of true positives.}
	\label{tab:link_pred_comparison_detailed}
	\centering
	\begin{tabular}{ lllll}
		\toprule
		\textbf{user} &\textbf{\#feedback}& \multicolumn{2}{c}{\textbf{true positives}}& \textbf{fraction}\\
		&&genetic & baseline& \\
		\midrule
		ex10 & 614 & 0.2822 & 0.1852 & 1.5420 \\
		ex2 & 675 & 0.2109 & 0.1259 & 1.6756 \\
		ex6 & 983 & 0.1913 & 0.0787 & 2.4312 \\
		ex5 & 1053 & 0.1554 & 0.0772 & 2.0123 \\
		ex14 & 1081 & 0.1837 & 0.1338 & 1.3727 \\
		ex13 & 1364 & 0.1987 & 0.1561 & 1.2728 \\
		ex4 & 2573 & 0.3392 & 0.1240 & 2.7350 \\
		ex3 & 2668 & 0.1458 & 0.0451 & 3.2336 \\
		ex11 & 4450 & 0.2154 & 0.0833 & 2.5867 \\
		ex7 & 6342 & 0.3533 & 0.1436 & 2.4599 \\
		ex9 & 7091 & 0.3247 & 0.1508 & 2.1536 \\
		\midrule
		\textbf{Average} & 2627 & 0.2364 & 0.1185 & 2.1325 \\
		\bottomrule
	\end{tabular}

\end{table}

Table \ref{tab:link_pred_comparison_detailed} depicts the true positives on user level. For all users, the genetic link prediction method outperforms the baseline heuristic.  On average, the link prediction method outperforms the baseline overall experiment participants by a factor of $2.13$.  Moreover, the figure also depicts the number of evaluated links per user. On average, a user evaluates 2627 links. Assuming a time of five seconds for a user to evaluate a link, users spent on average $3.6$ hours evaluating links which indicates, considering that the utilization of the feedback mechanism is not incentivized, that the feedback mechanism is practical.

\begin{figure}[!htb]
	
	\centering
	
	\subfigure[CDF of existing and non-existing links illustrating the attained larger similarity values of existing links when compared to non-existing links.]{
		\includegraphics[width=0.48\columnwidth]{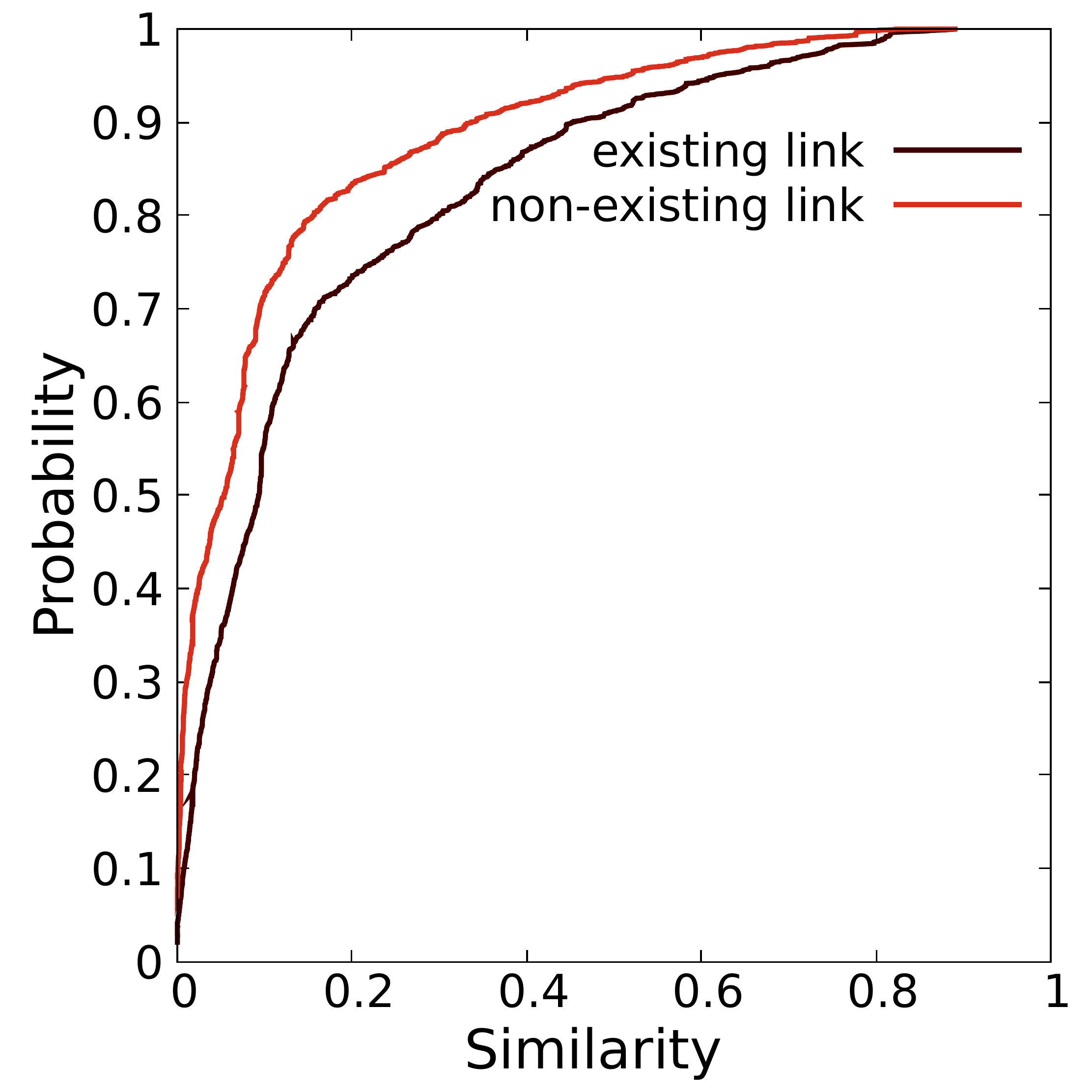}\label{fig:4_density_similarity_per_feedback}}
	\subfigure[Average time required to calculate the weights of Equation \ref{eq:linear_combined_similarity_metrics} by genetic programming algorithm (Algorithm \ref{eq:geneticProgramming})]{    \includegraphics[width=0.48\columnwidth]{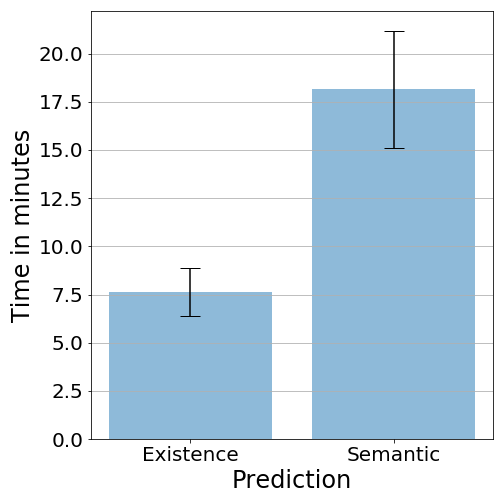}\label{fig:average_calculation_time}}
	
	\caption{}
	\label{fig:metric_top_error}
\end{figure}

Figure \ref{fig:4_density_similarity_per_feedback} illustrates the cumulative distribution function~(CDF) for similarity values of links that have been evaluated as existing~(1) and non-existing~(0) by the users. One notices, that an existing link has a higher probability of having a high similarity value than a non-existing link and that in turn, non-existing links have a higher probability for low similarity values. Thus, the knowledge builder assigns higher similarity values to existing links than to non-existing links which indicates that the knowledge builder distinguishes between existing and non-existing links.  

In the experiment, always the top-ranked results of the candidate set are recommended to the user. In particular, no threshold in the form of a specific similarity value is utilized that would prevent recommendations of links in case the candidate set consists entirely of non-existing links having low similarity values. In future work, such a threshold could be introduced to achieve higher true positives probabilities by removing those links automatically from the recommendations that have a low similarity value.

Figure \ref{fig:average_calculation_time} depicts the average time genetic programming requires to recalculate the weights for the existence and semantic prediction on users' phones (Equation \ref{eq:linear_combined_similarity_metrics}, Section \ref{sec:obtaining_weights}). The weight calculation for the existence prediction takes on average 7.5 minutes and for semantic 17.5 minutes. As the weights are recalculated on average every two hours to account for new user feedback, it is concluded that the deployment of the knowledge graph builder on users' phones is feasible.

\subsection{Metric weights}
\label{sec:eval_metric_weights}

\begin{figure}[!htb]
	\includegraphics[width=\columnwidth]{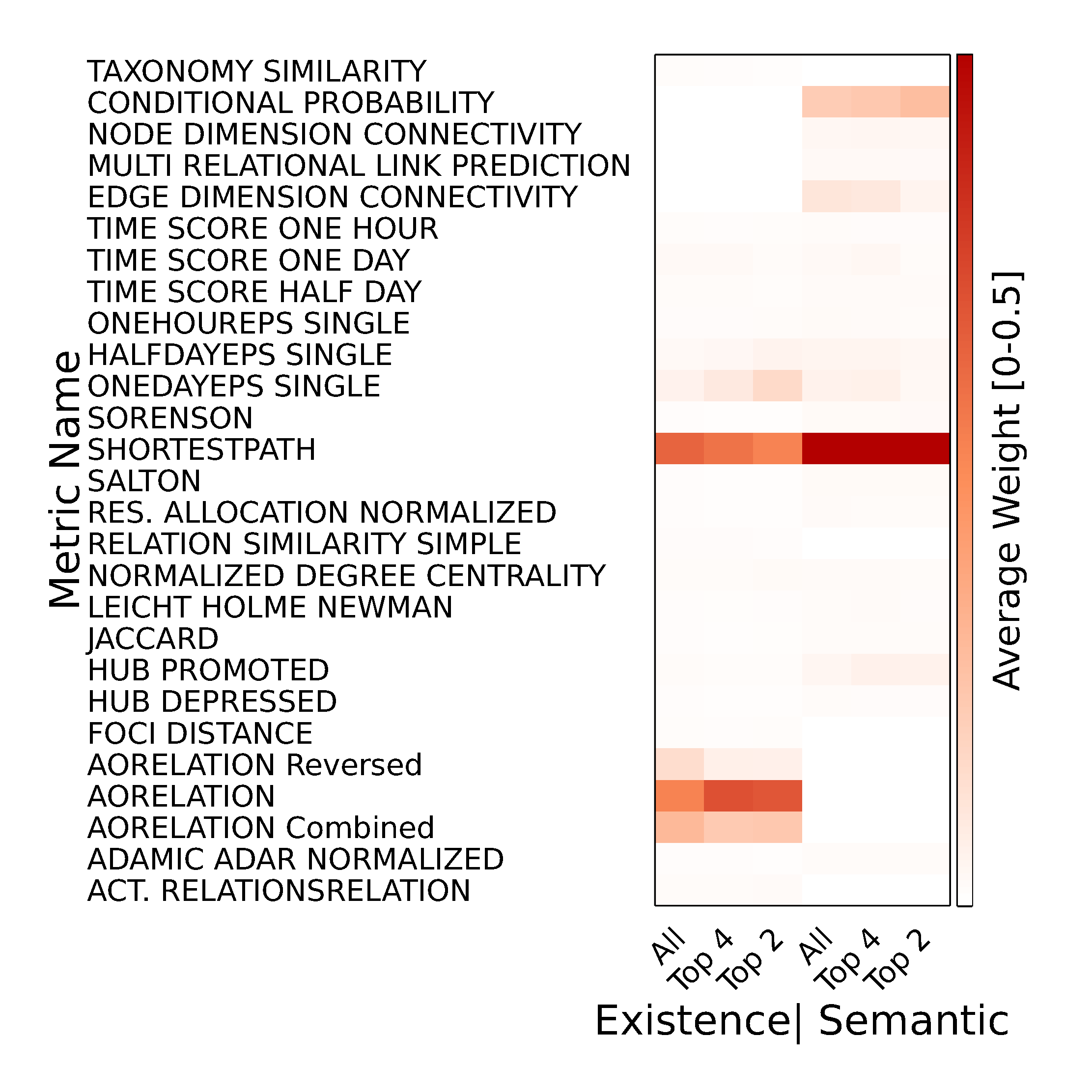}
	\caption{Weights of utilized similarity averaged over all experiment participants, the top 4 and top 2 participants by amount of collected feedback at the end of the experiment phase}
	\label{fig:heat_map_all_metric}
\end{figure}
Figure \ref{fig:heat_map_all_metric} depicts the utilized metrics, their weights averaged over all experiment participants, and by total amount of collected feedback the top 4 and top 2 participants. 

In the case of existence prediction, a combination of topological (Shortest path), semantical (AORelation), and temporal (OneDayEps) metrics are dominant to predict links. In semantic prediction, shortest path dominates the ensemble of deciding metrics. Nevertheless, also other topological metrics (hub promoted, Edge dimension connectivity), semantic (Conditional probability), and temporal (Oneday eps, half day eps) metrics contribute significantly to the predictions. In none of the scenarios (top 2, top 4, all users) a single metric is found to determine alone the link prediction. Nevertheless, in both types of predictions, the shortest path shows a large dominance when compared to the other metrics.
In particular, for both ensembles, a hierarchy can be observed in the dominance of the metric weights: Temporal metrics have lower weights. An explanation for this is that due to the short time period of the experiment (seven days) temporal patterns could not be sufficiently observed and learned. Moreover, in the case of existence prediction, an increase in the number of supervisions results in lower dominance of the shortest path and an increase in importance of semantic and temporal metrics. Both observations indicate that the algorithm is challenged by the cold start and that its performance could be improved when the study period is extended.

%
%
%

%
%
%

\section{Summary of findings}
\label{sec:summary_of_findings}
The key findings of the performed experiment are summarized as follow:
\begin{itemize}
	\item The knowledge graph builder is practical. In particular, its usability is indicated by the large number of supervision actions performed by users and its feasibility to run locally on users' phones.
	\item Optimizing the weighting of diverse similarity metrics for link prediction with genetic programming outperforms a baseline heuristic with regard to accuracy. In particular, links with a higher similarity value are recommended to the user which improves the acceptance of predicted links.
	\item An ensemble of semantic and temporal metrics are identified that dominate link prediction in a smart city application domain. This confirms findings from the literature that an ensemble of metrics can outperform a single metric.
	\item The novel metric AOrelation is dominant in the evaluated link existence prediction scenario.
\end{itemize}

In a nutshell, the findings demonstrate that the contributions of this paper support the domain-independent building of scalable and trustworthy knowledge graphs. 
In particular, the automation scales up the building process by suggesting links of high similarity accurately to users. The local human-supervision, which is considered as the gold standard in knowledge graph evaluation \cite{paulheim2017knowledge}, facilitates trust in the constructed knowledge graph. And the learning of underlying mechanisms that guide link formation in knowledge graph in the form of dominant metric weight ensembles indicates that the knowledge graph builder can be applied domain-independently to novel applications.

\section{Conclusion and Future Work}
\label{sec:conclusion}
This paper argues that an accurate and automated knowledge graph builder for cyber-physical-social systems can be constructed that accounts for values such as privacy-preservation and accountability. By applying a value-sensitive design approach 
a system is designed that builds knowledge graphs automatically while remaining accountable to humans via local human supervision and thus can be applied effectively to novel application domains such as smart cities. In particular, localized supervision is considered as the gold standard in knowledge graph evaluation and thus increases the trust in the constructed graph, while the automation facilitates the scalability of the building process.
This is evaluated by a methodology that integrates the constructed system into users' daily lives.

The results point to various avenues for future research. First, the identification of dominant similarity metrics in a smart city application scenario suggests to further investigate these metrics in varying application domains. In particular, the domain-independence of the knowledge graph builder could be further demonstrated. 
Second, several machine learning models utilized in automation are not explainable~\cite{ribeiro2016should} which limits users' trust \cite{friedman1997software}. The transparent display of metric weights in the metric dashboard of the knowledge graph builder could be a basis for the explainability of recommended links. In particular, a user could reason why a link was recommended based on the observed metric weights. Third, the participant field of the user study and the time frame of the experiment could be enlarged to add significance to the identified findings, reduce the cold start problem and identify further temporal patterns which could improve the recommendation accuracy. Finally, the parameters of the knowledge graph builder could be fine-tuned by a meta-optimization strategy to improve the prediction accuracy.

\appendix
%

\section*{Genetic Programming}
\label{ap:training}

Genetic programming (Section \ref{sec:genetic_algorithm}) is utilized to calculate the weights $a_i$ of Equation \ref{eq:linear_combined_similarity_metrics}. 
One way of implementing a genetic programming algorithm is described in Algorithm \ref{eq:geneticProgramming}. In the following, this implementation is illustrated in greater detail. The reader is referred to Koza \cite{koza:genetic} for definitions and motivations of utilized terms.

\paragraph{Fitness Evaluation Function}    The function is defined from the space of weight vectors $\vec{a}$  and training sets $I$ (Section \ref{sec:obtaining_weights}) to the space of real numbers: $f(R^n,I) \rightarrow R$, 
where $n$ is the number of utilized similarity measures. 
A training set with positive and negative examples of links between nodes is used to calculate how close a particular $\vec{a}$ predicts the existence of a link, respectively type of a link (Section \ref{sec:problem_formulation}).

Let $I$ be a concrete set of training instances, having the size~$m$. Let $i\in I$ be a single training instance, then in case of link existence prediction the instance $i = (u,v,\{0,1\})$ is an array of size three (Section \ref{sec:obtaining_weights}), $u$ being the target and $v$ the candidate node. 
Then a mean squared error evaluation is utilized as the fitness function:

\begin{equation}
\begin{aligned}
f(\vec{a},I):= \text{MSE($\vec{a}$,I)}  &= \frac{1}{m}\sum_{i \in I} \big(s(i[0],i[1])-i[2] \big)^2 \\
&= \frac{1}{m}\sum_{i \in I} \Big(\big(\sum_{i} a_i s_i(i[0],i[1])\big) - i[2] \Big)^2  
\end{aligned}
\end{equation}

\paragraph{Genotype and phenotype}
The genotype of an individual is its weight vector $\vec{a}$. This vector is stored in a linked list, where the last element points to the first. 
The phenotype of an individual is then its fitness value, calculated via its weight vector and the training instances by $f(a, I)$. The smaller the fitness value of an individual, the better is the genotype (weight vector $\vec{a}$) of that individual able to predict links in the training set.

\paragraph{Crossover} A single-point crossover is chosen. 
The linked list of the weight vector $\vec{a}$ is split randomly at the same position for both parents and then two children are created. 
\paragraph{Mutation}
Mutation is simulated by identifying randomly a position in the weight vector $\vec{a}$ and altering its value if a specific threshold is matched. 
The new value is randomly chosen from a uniform distribution in the interval~$[0,1)$.
\paragraph{Selection}
The least fit individual of a generation is removed from reproduction.
\paragraph{Micro approach}
A micro genetic programming approach is utilized by considering small population sizes with 5 to 11 individuals per generation. Please refer to Hafner \cite{hafner:post} for details.

\begin{algorithm}
	\caption{Genetic Programming}\label{eq:geneticProgramming}
	\begin{algorithmic}[1]
		\Procedure{run}{}\Comment{obtain weights for similarity measures}
		\State initialisePopulation()
		\State $\text{notDone} \gets \text{TRUE}$
		\State iter $\gets 0$
		\While{$\text{notDone} \;\& \& \; (\text{iter}<\text{maxIter})$} 
		\State iter++
		\State population.run()
		\State tmpBestIndivid $\gets $ population.getBest()
		\If {tmpBestIndivid.getFitness() $<$ bestIndivid.getFitness()}
		\State bestIndividual $ \gets$ tmpBestIndividual
		\If {bestIndividual.getFitness() $<$ tol}
		\State notDone $\gets$ FALSE
		\EndIf
		\EndIf
		population.purge()
		\EndWhile
		\State \textbf{return} bestIndivid.getGenotype()  \Comment{return weight vector $\vec{a}$}
		\EndProcedure
	\end{algorithmic}
\end{algorithm}


\section*{Novel and adjusted similarity measures}
\label{sec:algo_utilized_similarity_measures}
Table \ref{tab:classical_metrics} illustrates the similarity measures utilized in the experiment. Some of those are obtained by modifying metrics found in literature or are introduced in this work.
These two types of metrics will be described in the following.

\subsection{Modified Metrics}

\begin{itemize}
	\item \textbf{Adamic Adar (AA)}    
	In order to normalize the measure onto the interval $[0,1]$, a scaling term is introduced. 
	
	\begin{equation}
	A(u,v) = \Bigg(\sum_{z\in \Gamma (u) \cap \Gamma(v)} \frac{1}{log(|\Gamma(z)|)} \Bigg) \frac{1}{|\Gamma(u)\cap \Gamma(v)|\frac{1}{log(2)}}
	\end{equation}


	\item \textbf{Resource Allocation (R)}
	This measure is also normalized:
	\begin{equation}
	R(u,v) = \Bigg( \sum_{z \in \Gamma(u) \cap \Gamma(v)} \frac{1}{|\Gamma(z)|} \Bigg) \frac{1}{|\Gamma(u)\cap \Gamma(v)|\frac{1}{2}}
	\end{equation}

	\item \textbf{Focci distance} The measure found in Jahanbakhsh et. al \cite{jahanbakhsh2012predicting} is adjusted because a neighborhood cannot be defined in the same way as in the referenced work.
	In this work, a neighborhood is defined as the union of nodes to which both nodes $u$ and $v$ are connected via the same relation identifier. Thus the Focci distance looks as follow
	
	\begin{equation}
	FD(u,v) = \underset{j\in J(u)\cap J(v)}{max} \Bigg(\underset{z\in \Gamma(u,j)\cap \Gamma(v,j)}{max} \frac{1}{|\Gamma(z,\text{inverse}(j))|}\Bigg)
	\end{equation}
	

	\item \textbf{Shortest Path (SP)} Due to the computational complexity of path-based methods, this measure is restricted to paths of maximum length 5. If a shortest path is longer than $5$, than the similarity value is $0$:
	
	\begin{equation}
	\begin{aligned}
	SP(u,v) = max \Big(0, 1- \frac{\underset{n}{min}P_n(u,v)-1}{5}\Big)\\
	\end{aligned}
	\end{equation}

	\item \textbf{Time Score (TS)} A modified version of the Time Score metric, as introduced in Munasinghe     \cite{munasinghe2012time}, is utilized:
	The adjusted version accounts for times given in milliseconds, not time steps. This is accomplished by dividing the time given in milliseconds by the discouting time. I.e. if a time step should have the length of one day than the equivalent amount of milliseconds is taken as a discounting factor. Moreover, TS is normalized to $[0,1]$ by discouting with $d$, which normalizes the harmonic mean of co-occurrences:
	\begin{equation}
	\begin{aligned}
	&TS(u,v) &&= \sum_{c_i \in M} \frac{H_m(u,v,c_i) d(u,v,c) \beta^{k(u,v,c)}}{floor(|t(u,c) - t(v,c)|/mm)+1}\\
	& M  && =\Gamma(u) \cap \Gamma(v)\\
	& t(u,c) && \text{most recent time stamp of realized link}\\
	&&& \text{ between $u$ and $c$, given in miliseconds}\\
	&mm && \text{miliseconds to discount. Discounts}\\
	&&&\text{continuous time to time steps}\\
	& k(u,v,c) && = floor\Big(\frac{min(t(u,c), t(v,c))}{mm}\Big) \\
	& H_m^i(u_1,u_2,c) &&= \frac{1}{\frac{1}{2}\sum_{i=1}^2\frac{1}{|J(c,u_i)|}} \\
	&d(u,v,c) = &&\frac{1}{max(|J(u,c)|,|J(v,c)|)},
	\end{aligned}
	\end{equation}
\end{itemize}

\subsection{Novel Metrics}
\label{sec:new_metrics}
Two novel semantic metrics for are introduced that are independent of the topological distance of the target and candidate node.
These novel metrics are:

\begin{itemize}
	\item \textbf{Active Relations Relation (ARR)}
	This measure is the Jaccard index for the active relationships of two nodes. Hence it counts the number of active relationships in which both, target node $u$ and candidate node $v$ engage and divides the result by the number of all relationships in which $u$ engages. 
	\begin{equation}
	ARR(u,v) = \frac{J(u) \cap J(v)}{J(u)}
	\end{equation}
	The idea is, that two nodes are more similar when they share the same type of relationships in which they actively engage.
	
	\item \textbf{AO Relation (AOR)}
	Counts the number of neighbors of candidate node $v$ which are of the same concept as the target node $u$ and divides the number by the amount of neighbors of $v$:
	\begin{equation}
	AOR(u,v) = \frac{\sum_{z\in \Gamma(v)} Eq(\text{type}(u), \text{type}(z))}{|\Gamma(v)|}
	\end{equation}
	where $Eq$ returns $1$ if both types are equal. Else it returns $0$.
	The rationale is, that a relationship will form more likely if the candidate node $v$ already engages in relationships with nodes of the same type as $u$.
	This metric $m(u,v)$ has a reversed version $mr(u,v)$ which is defined as: $mr(u,v):=m(v,u)$ and a combined version $mc(u,v)$, which is defined as $mc(u,v):= \frac{m(u,v)+mr(u,v)}{2}$.
\end{itemize}



\bibliographystyle{myIEEEtran}
\bibliography{IEEEabrv,refs}

\begin{thebibliography}{10}
\providecommand{\url}[1]{#1}
\csname url@samestyle\endcsname
\providecommand{\newblock}{\relax}
\providecommand{\bibinfo}[2]{#2}
\providecommand{\BIBentrySTDinterwordspacing}{\spaceskip=0pt\relax}
\providecommand{\BIBentryALTinterwordstretchfactor}{4}
\providecommand{\BIBentryALTinterwordspacing}{\spaceskip=\fontdimen2\font plus
\BIBentryALTinterwordstretchfactor\fontdimen3\font minus
  \fontdimen4\font\relax}
\providecommand{\BIBforeignlanguage}[2]{{%
\expandafter\ifx\csname l@#1\endcsname\relax
\typeout{** WARNING: IEEEtran.bst: No hyphenation pattern has been}%
\typeout{** loaded for the language `#1'. Using the pattern for}%
\typeout{** the default language instead.}%
\else
\language=\csname l@#1\endcsname
\fi
#2}}
\providecommand{\BIBdecl}{\relax}
\BIBdecl

\bibitem{chakraborty2016automotive}
S.~Chakraborty, M.~A. Al~Faruque, W.~Chang, D.~Goswami, M.~Wolf, and Q.~Zhu,
  ``Automotive cyber--physical systems: A tutorial introduction,'' \emph{IEEE
  Design \& Test}, vol.~33, no.~4, pp. 92--108, 2016.

\bibitem{dautov2018data}
R.~Dautov, S.~Distefano, D.~Bruneo, F.~Longo, G.~Merlino, and A.~Puliafito,
  ``Data processing in cyber-physical-social systems through edge computing,''
  \emph{IEEE Access}, vol.~6, pp. 29\,822--29\,835, 2018.

\bibitem{zhang2018cyber}
J.~J. Zhang, F.-Y. Wang, X.~Wang, G.~Xiong, F.~Zhu, Y.~Lv, J.~Hou, S.~Han,
  Y.~Yuan, Q.~Lu \emph{et~al.}, ``Cyber-physical-social systems: The state of
  the art and perspectives,'' \emph{IEEE Transactions on Computational Social
  Systems}, vol.~5, no.~3, pp. 829--840, 2018.

\bibitem{dudas2009onalin}
P.~M. Dudas, M.~Ghafourian, and H.~A. Karimi, ``Onalin: Ontology and algorithm
  for indoor routing,'' in \emph{2009 Tenth International Conference on Mobile
  Data Management: Systems, Services and Middleware}.\hskip 1em plus 0.5em
  minus 0.4em\relax IEEE, 2009, pp. 720--725.

\bibitem{hu2016personal}
H.~Hu, A.~Elkus, and L.~Kerschberg, ``A personal health recommender system
  incorporating personal health records, modular ontologies, and crowd-sourced
  data,'' in \emph{2016 IEEE/ACM International Conference on Advances in Social
  Networks Analysis and Mining (ASONAM)}.\hskip 1em plus 0.5em minus
  0.4em\relax IEEE, 2016, pp. 1027--1033.

\bibitem{wiesner2010adapting}
M.~Wiesner and D.~Pfeifer, ``Adapting recommender systems to the requirements
  of personal health record systems,'' in \emph{Proceedings of the 1st ACM
  International Health Informatics Symposium}, 2010, pp. 410--414.

\bibitem{hao2017end}
Y.~Hao, Y.~Zhang, K.~Liu, S.~He, Z.~Liu, H.~Wu, and J.~Zhao, ``An end-to-end
  model for question answering over knowledge base with cross-attention
  combining global knowledge,'' in \emph{Proceedings of the 55th Annual Meeting
  of the Association for Computational Linguistics (Volume 1: Long Papers)},
  2017, pp. 221--231.

\bibitem{caragea2009ontology}
D.~Caragea, V.~Bahirwani, W.~Aljandal, and W.~H. Hsu, ``Ontology-based link
  prediction in the livejournal social network,'' in \emph{Eighth Symposium on
  Abstraction, Reformulation, and Approximation}, 2009.

\bibitem{cui2019infer}
Z.~Cui, L.~Pan, S.~Liu, and L.~Cui, ``Infer latent privacy for attribute
  network in knowledge graph,'' in \emph{2019 IEEE International Conference on
  Big Data (Big Data)}.\hskip 1em plus 0.5em minus 0.4em\relax IEEE, 2019, pp.
  2542--2551.

\bibitem{alani2003automatic}
H.~Alani, S.~Kim, D.~E. Millard, M.~J. Weal, W.~Hall, P.~H. Lewis, and N.~R.
  Shadbolt, ``Automatic ontology-based knowledge extraction from web
  documents,'' \emph{IEEE Intelligent Systems}, vol.~18, no.~1, pp. 14--21,
  2003.

\bibitem{wang2013boosting}
Z.~Wang, J.~Li, and J.~Tang, ``Boosting cross-lingual knowledge linking via
  concept annotation,'' in \emph{Twenty-Third International Joint Conference on
  Artificial Intelligence}, 2013.

\bibitem{schafer2017towards}
H.~Sch{\"a}fer, S.~Hors-Fraile, R.~P. Karumur, A.~Calero~Valdez, A.~Said,
  H.~Torkamaan, T.~Ulmer, and C.~Trattner, ``Towards health (aware) recommender
  systems,'' in \emph{Proceedings of the 2017 international conference on
  digital health}, 2017, pp. 157--161.

\bibitem{dussell1999position}
W.~O. Dussell, J.~M. Janky, J.~F. Schipper, and D.~J. Cowl, ``Position based
  personal digital assistant,'' Aug.~17 1999, uS Patent 5,938,721.

\bibitem{asikis2020value}
T.~Asikis, J.~Klinglmayr, D.~Helbing, and E.~Pournaras, ``How value-sensitive
  design can empower sustainable consumption,'' \emph{arXiv preprint
  arXiv:2004.09180}, 2020.

\bibitem{zhou2011effect}
T.~Zhou, ``The effect of initial trust on user adoption of mobile payment,''
  \emph{Information Development}, vol.~27, no.~4, pp. 290--300, 2011.

\bibitem{belanger2008trust}
F.~B{\'e}langer and L.~Carter, ``Trust and risk in e-government adoption,''
  \emph{The Journal of Strategic Information Systems}, vol.~17, no.~2, pp.
  165--176, 2008.

\bibitem{nahavandi2017trusted}
S.~Nahavandi, ``Trusted autonomy between humans and robots: Toward
  human-on-the-loop in robotics and autonomous systems,'' \emph{IEEE Systems,
  Man, and Cybernetics Magazine}, vol.~3, no.~1, pp. 10--17, 2017.

\bibitem{friedman1997software}
B.~Friedman and H.~Nissenbaum, ``Software agents and user autonomy,'' in
  \emph{Proceedings of the first international conference on Autonomous
  agents}, 1997, pp. 466--469.

\bibitem{van2008moral}
J.~Van~den Hoven, ``Moral methodology and information technology,'' \emph{The
  handbook of information and computer ethics}, vol.~49, 2008.

\bibitem{van2010use}
------, ``The use of normative theories in computer ethics,'' \emph{The
  Cambridge handbook of information and computer ethics}, pp. 59--76, 2010.

\bibitem{hill1991autonomy}
T.~E. Hill~Jr, \emph{Autonomy and self-respect}.\hskip 1em plus 0.5em minus
  0.4em\relax Cambridge University Press, 1991.

\bibitem{friedman2003human}
B.~Friedman and P.~H. Kahn~Jr, ``Human values, ethics, and design,'' \emph{The
  human-computer interaction handbook}, pp. 1177--1201, 2003.

\bibitem{bliss2014evolutionary}
C.~A. Bliss, M.~R. Frank, C.~M. Danforth, and P.~S. Dodds, ``An evolutionary
  algorithm approach to link prediction in dynamic social networks,''
  \emph{Journal of Computational Science}, vol.~5, no.~5, pp. 750--764, 2014.

\bibitem{pazzani1999framework}
M.~J. Pazzani, ``A framework for collaborative, content-based and demographic
  filtering,'' \emph{Artificial intelligence review}, vol.~13, no. 5-6, pp.
  393--408, 1999.

\bibitem{van2000using}
R.~Van~Meteren and M.~Van~Someren, ``Using content-based filtering for
  recommendation,'' in \emph{Proceedings of the Machine Learning in the New
  Information Age: MLnet/ECML2000 Workshop}, vol.~30, 2000, pp. 47--56.

\bibitem{basilico2004unifying}
J.~Basilico and T.~Hofmann, ``Unifying collaborative and content-based
  filtering,'' in \emph{Proceedings of the twenty-first international
  conference on Machine learning}, 2004, p.~9.

\bibitem{isinkaye2015recommendation}
F.~Isinkaye, Y.~Folajimi, and B.~Ojokoh, ``Recommendation systems: Principles,
  methods and evaluation,'' \emph{Egyptian Informatics Journal}, vol.~16,
  no.~3, pp. 261--273, 2015.

\bibitem{friedman2015privacy}
A.~Friedman, B.~P. Knijnenburg, K.~Vanhecke, L.~Martens, and S.~Berkovsky,
  ``Privacy aspects of recommender systems,'' in \emph{Recommender Systems
  Handbook}.\hskip 1em plus 0.5em minus 0.4em\relax Springer, 2015, pp.
  649--688.

\bibitem{carlier2011combining}
A.~Carlier, G.~Ravindra, V.~Charvillat, and W.~T. Ooi, ``Combining
  content-based analysis and crowdsourcing to improve user interaction with
  zoomable video,'' in \emph{Proceedings of the 19th ACM international
  conference on Multimedia}, 2011, pp. 43--52.

\bibitem{goeau2011visual}
H.~Go{\"e}au, A.~Joly, S.~Selmi, P.~Bonnet, E.~Mouysset, L.~Joyeux, J.-F.
  Molino, P.~Birnbaum, D.~Bathelemy, and N.~Boujemaa, ``Visual-based plant
  species identification from crowdsourced data,'' in \emph{Proceedings of the
  19th ACM international conference on Multimedia}, 2011, pp. 813--814.

\bibitem{ferman2002content}
A.~M. Ferman, J.~H. Errico, P.~v. Beek, and M.~I. Sezan, ``Content-based
  filtering and personalization using structured metadata,'' in
  \emph{Proceedings of the 2nd ACM/IEEE-CS joint conference on Digital
  libraries}, 2002, pp. 393--393.

\bibitem{yu2014personalized}
X.~Yu, X.~Ren, Y.~Sun, Q.~Gu, B.~Sturt, U.~Khandelwal, B.~Norick, and J.~Han,
  ``Personalized entity recommendation: A heterogeneous information network
  approach,'' in \emph{Proceedings of the 7th ACM international conference on
  Web search and data mining}, 2014, pp. 283--292.

\bibitem{shi2016discriminative}
B.~Shi and T.~Weninger, ``Discriminative predicate path mining for fact
  checking in knowledge graphs,'' \emph{Knowledge-based systems}, vol. 104, pp.
  123--133, 2016.

\bibitem{melo2017approach}
A.~Melo and H.~Paulheim, ``An approach to correction of erroneous links in
  knowledge graphs,'' in \emph{CEUR Workshop Proceedings}, vol. 2065.\hskip 1em
  plus 0.5em minus 0.4em\relax RWTH, 2017, pp. 54--57.

\bibitem{nickel2015review}
M.~Nickel, K.~Murphy, V.~Tresp, and E.~Gabrilovich, ``A review of relational
  machine learning for knowledge graphs,'' \emph{Proceedings of the IEEE}, vol.
  104, no.~1, pp. 11--33, 2015.

\bibitem{wang2019research}
B.~Wang, J.~Luo, and S.~Zhu, ``Research on domain ontology automation
  construction based on chinese texts,'' in \emph{Proceedings of the 2019 8th
  International Conference on Software and Computer Applications}, 2019, pp.
  425--430.

\bibitem{west2014knowledge}
R.~West, E.~Gabrilovich, K.~Murphy, S.~Sun, R.~Gupta, and D.~Lin, ``Knowledge
  base completion via search-based question answering,'' in \emph{Proceedings
  of the 23rd international conference on World wide web}, 2014, pp. 515--526.

\bibitem{suh2009singularity}
B.~Suh, G.~Convertino, E.~H. Chi, and P.~Pirolli, ``The singularity is not
  near: slowing growth of wikipedia,'' in \emph{Proceedings of the 5th
  International Symposium on Wikis and Open Collaboration}, 2009, pp. 1--10.

\bibitem{paulheim2017knowledge}
H.~Paulheim, ``Knowledge graph refinement: A survey of approaches and
  evaluation methods,'' \emph{Semantic web}, vol.~8, no.~3, pp. 489--508, 2017.

\bibitem{amador2019ontology}
E.~Amador-Dom{\'\i}nguez, P.~Hohenecker, T.~Lukasiewicz, D.~Manrique, and
  E.~Serrano, ``An ontology-based deep learning approach for knowledge graph
  completion with fresh entities,'' in \emph{International Symposium on
  Distributed Computing and Artificial Intelligence}.\hskip 1em plus 0.5em
  minus 0.4em\relax Springer, 2019, pp. 125--133.

\bibitem{chen2016fast}
B.~Chen, L.~Chen, and B.~Li, ``A fast algorithm for predicting links to nodes
  of interest,'' \emph{Information Sciences}, vol. 329, pp. 552--567, 2016.

\bibitem{haghani2019systemic}
S.~Haghani and M.~R. Keyvanpour, ``A systemic analysis of link prediction in
  social network,'' \emph{Artificial Intelligence Review}, vol.~52, no.~3, pp.
  1961--1995, 2019.

\bibitem{lichtenwalter2010new}
R.~N. Lichtenwalter, J.~T. Lussier, and N.~V. Chawla, ``New perspectives and
  methods in link prediction,'' in \emph{Proceedings of the 16th ACM SIGKDD
  international conference on Knowledge discovery and data mining}, 2010, pp.
  243--252.

\bibitem{martinez2016survey}
V.~Mart{\'\i}nez, F.~Berzal, and J.-C. Cubero, ``A survey of link prediction in
  complex networks,'' \emph{ACM Computing Surveys (CSUR)}, vol.~49, no.~4, pp.
  1--33, 2016.

\bibitem{cao2016link}
X.~Cao, Y.~Zheng, C.~Shi, J.~Li, and B.~Wu, ``Link prediction in schema-rich
  heterogeneous information network,'' in \emph{Pacific-Asia Conference on
  Knowledge Discovery and Data Mining}.\hskip 1em plus 0.5em minus 0.4em\relax
  Springer, 2016, pp. 449--460.

\bibitem{friemel:dynamics_of_social_networks}
T.~N. Friemel, ``Dissolution point and isolation robustness: robustness
  criteria for general cluster analysis methods,'' \emph{Procedia - Social and
  Behavioral Sciences}, vol.~22, pp. 2--3, 2011.

\bibitem{wang2015link}
P.~Wang, B.~Xu, Y.~Wu, and X.~Zhou, ``Link prediction in social networks: the
  state-of-the-art,'' \emph{Science China Information Sciences}, vol.~58,
  no.~1, pp. 1--38, 2015.

\bibitem{davis2013supervised}
D.~Davis, R.~Lichtenwalter, and N.~V. Chawla, ``Supervised methods for
  multi-relational link prediction,'' \emph{Social network analysis and
  mining}, vol.~3, no.~2, pp. 127--141, 2013.

\bibitem{shi2017survey}
C.~Shi, Y.~Li, J.~Zhang, Y.~Sun, and S.~Y. Philip, ``A survey of heterogeneous
  information network analysis,'' \emph{IEEE Transactions on Knowledge and Data
  Engineering}, vol.~29, no.~1, pp. 17--37, 2017.

\bibitem{tylenda2009towards}
T.~Tylenda, R.~Angelova, and S.~Bedathur, ``Towards time-aware link prediction
  in evolving social networks,'' in \emph{Proceedings of the 3rd workshop on
  social network mining and analysis}.\hskip 1em plus 0.5em minus 0.4em\relax
  ACM, 2009, p.~9.

\bibitem{yang2012predicting}
Y.~Yang, N.~Chawla, Y.~Sun, and J.~Hani, ``Predicting links in multi-relational
  and heterogeneous networks,'' in \emph{Data Mining (ICDM), 2012 IEEE 12th
  International Conference on}.\hskip 1em plus 0.5em minus 0.4em\relax IEEE,
  2012, pp. 755--764.

\bibitem{maedche2002clustering}
A.~Maedche and V.~Zacharias, ``Clustering ontology-based metadata in the
  semantic web,'' in \emph{PKDD}, vol.~2.\hskip 1em plus 0.5em minus
  0.4em\relax Springer, 2002, pp. 348--360.

\bibitem{grimnes2008instance}
G.~A. Grimnes, P.~Edwards, and A.~Preece, ``Instance based clustering of
  semantic web resources,'' in \emph{European Semantic Web Conference}.\hskip
  1em plus 0.5em minus 0.4em\relax Springer, 2008, pp. 303--317.

\bibitem{opuszko2012classification}
M.~Opuszko and J.~Ruhland, ``Classification analysis in complex online social
  networks using semantic web technologies,'' in \emph{Advances in Social
  Networks Analysis and Mining (ASONAM), 2012 IEEE/ACM International Conference
  on}.\hskip 1em plus 0.5em minus 0.4em\relax IEEE, 2012, pp. 1032--1039.

\bibitem{ma2019jointly}
W.~Ma, M.~Zhang, Y.~Cao, W.~Jin, C.~Wang, Y.~Liu, S.~Ma, and X.~Ren, ``Jointly
  learning explainable rules for recommendation with knowledge graph,'' in
  \emph{The World Wide Web Conference}, 2019, pp. 1210--1221.

\bibitem{brandao2013using}
M.~A. Brand{\~a}o, M.~M. Moro, G.~R. Lopes, and J.~P. Oliveira, ``Using link
  semantics to recommend collaborations in academic social networks,'' in
  \emph{Proceedings of the 22nd International Conference on World Wide
  Web}.\hskip 1em plus 0.5em minus 0.4em\relax ACM, 2013, pp. 833--840.

\bibitem{ozcan2019multivariate}
A.~Ozcan and S.~G. Oguducu, ``Multivariate time series link prediction for
  evolving heterogeneous network,'' \emph{International Journal of Information
  Technology \& Decision Making}, vol.~18, no.~01, pp. 241--286, 2019.

\bibitem{vani2015investigating}
K.~Vani and D.~Gupta, ``Investigating the impact of combined similarity metrics
  and pos tagging in extrinsic text plagiarism detection system,'' in
  \emph{2015 International Conference on Advances in Computing, Communications
  and Informatics (ICACCI)}.\hskip 1em plus 0.5em minus 0.4em\relax IEEE, 2015,
  pp. 1578--1584.

\bibitem{chen:similarity_metric}
S.~Chen, B.~Ma, and K.~Zhang, ``On the similarity metric and the distance
  metric,'' \emph{Theoretical Computer Science}, vol. 410, no. 24-25, pp.
  2365--2376, 2009.

\bibitem{koza:genetic}
J.~R. Koza, \emph{Genetic Programming}.\hskip 1em plus 0.5em minus 0.4em\relax
  Cambridge, MA: The MIT Press, 1992.

\bibitem{sastry2019survey}
H.~G. Sastry, L.~C. Reddy \emph{et~al.}, ``A survey on genetic programming in
  data mining tasks,'' \emph{Journal of Computer Technology \& Applications},
  vol.~3, no.~1, pp. 9--15, 2019.

\bibitem{espejo2009survey}
P.~G. Espejo, S.~Ventura, and F.~Herrera, ``A survey on the application of
  genetic programming to classification,'' \emph{IEEE Transactions on Systems,
  Man, and Cybernetics, Part C (Applications and Reviews)}, vol.~40, no.~2, pp.
  121--144, 2009.

\bibitem{petke2017genetic}
J.~Petke, S.~O. Haraldsson, M.~Harman, W.~B. Langdon, D.~R. White, and J.~R.
  Woodward, ``Genetic improvement of software: a comprehensive survey,''
  \emph{IEEE Transactions on Evolutionary Computation}, vol.~22, no.~3, pp.
  415--432, 2017.

\bibitem{khan2019recent}
A.~Khan, A.~S. Qureshi, N.~Wahab, M.~Hussain, and M.~Y. Hamza, ``A recent
  survey on the applications of genetic programming in image processing,''
  \emph{arXiv preprint arXiv:1901.07387}, 2019.

\bibitem{nguyen2017genetic}
S.~Nguyen, Y.~Mei, and M.~Zhang, ``Genetic programming for production
  scheduling: a survey with a unified framework,'' \emph{Complex \& Intelligent
  Systems}, vol.~3, no.~1, pp. 41--66, 2017.

\bibitem{agapitos2019survey}
A.~Agapitos, R.~Loughran, M.~Nicolau, S.~Lucas, M.~O’Neill, and A.~Brabazon,
  ``A survey of statistical machine learning elements in genetic programming,''
  \emph{IEEE Transactions on Evolutionary Computation}, vol.~23, no.~6, pp.
  1029--1048, 2019.

\bibitem{jaccard1901bulletin}
P.~Jaccard, ``Bulletin de la soci{\'e}t{\'e} vaudoise des sciences
  naturelles,'' \emph{Etude comparative de la distribution florale dans une
  portion des Alpes et des Jura}, vol.~37, pp. 547--579, 1901.

\bibitem{adamic2003friends}
L.~A. Adamic and E.~Adar, ``Friends and neighbors on the web,'' \emph{Social
  networks}, vol.~25, no.~3, pp. 211--230, 2003.

\bibitem{zhou2009predicting}
T.~Zhou, L.~L{\"u}, and Y.-C. Zhang, ``Predicting missing links via local
  information,'' \emph{The European Physical Journal B}, vol.~71, no.~4, pp.
  623--630, 2009.

\bibitem{ravasz2002hierarchical}
E.~Ravasz, A.~L. Somera, D.~A. Mongru, Z.~N. Oltvai, and A.-L. Barab{\'a}si,
  ``Hierarchical organization of modularity in metabolic networks,''
  \emph{science}, vol. 297, no. 5586, pp. 1551--1555, 2002.

\bibitem{leicht2006vertex}
E.~A. Leicht, P.~Holme, and M.~E. Newman, ``Vertex similarity in networks,''
  \emph{Physical Review E}, vol.~73, no.~2, p. 026120, 2006.

\bibitem{mcgill1983introduction}
M.~McGill and G.~Salton, ``Introduction to modern information retrieval.
  1983,'' \emph{McGraw-Hil, New York}, 1983.

\bibitem{sorensen1948method}
T.~S{\o}rensen, T.~S{\o}rensen, T.~S{\o}rensen, T.~SORENSEN, T.~Sorensen,
  T.~Sorensen, and T.~Biering-S{\o}rensen, ``A method of establishing groups of
  equal amplitude in plant sociology based on similarity of species content and
  its application to analyses of the vegetation on danish commons,'' 1948.

\bibitem{munasinghe2012time}
L.~Munasinghe and R.~Ichise, ``Time score: A new feature for link prediction in
  social networks,'' \emph{IEICE TRANSACTIONS on Information and Systems},
  vol.~95, no.~3, pp. 821--828, 2012.

\bibitem{jahanbakhsh2012predicting}
K.~Jahanbakhsh, V.~King, and G.~C. Shoja, ``Predicting missing contacts in
  mobile social networks,'' \emph{Pervasive and Mobile Computing}, vol.~8,
  no.~5, pp. 698--716, 2012.

\bibitem{rossetti2011scalable}
G.~Rossetti, M.~Berlingerio, and F.~Giannotti, ``Scalable link prediction on
  multidimensional networks,'' in \emph{Data Mining Workshops (ICDMW), 2011
  IEEE 11th International Conference on}.\hskip 1em plus 0.5em minus
  0.4em\relax IEEE, 2011, pp. 979--986.

\bibitem{crisp1987persuasive}
R.~Crisp, ``Persuasive advertising, autonomy, and the creation of desire,''
  \emph{Journal of Business Ethics}, vol.~6, no.~5, pp. 413--418, 1987.

\bibitem{mei2011power}
S.~Mei, X.~Zhang, and M.~Cao, \emph{Power grid complexity}.\hskip 1em plus
  0.5em minus 0.4em\relax Springer Science \& Business Media, 2011.

\bibitem{ribeiro2016should}
M.~T. Ribeiro, S.~Singh, and C.~Guestrin, ``" why should i trust you?"
  explaining the predictions of any classifier,'' in \emph{Proceedings of the
  22nd ACM SIGKDD international conference on knowledge discovery and data
  mining}, 2016, pp. 1135--1144.

\bibitem{hafner:post}
C.~Hafner, \emph{Post-Modern Electromagnetics Using Intelligent MaXwell
  Solvers}.\hskip 1em plus 0.5em minus 0.4em\relax Chichester: John Wiley \&
  Sons, 1999.

\end{thebibliography}
%



%
%

\end{document}